# The early Earth as an analogue for exoplanetary biogeochemistry

Eva E. Stüeken[1,*], Stephanie L. Olson[2], Eli Moore[3], Bradford J. Foley[4]

## Abstract

Planet Earth has evolved over the past 4.5 billion years from an entirely anoxic planet with possibly a different tectonic regime to the oxygenated world with horizontal plate tectonics that we know today. For most of this time, Earth has been inhabited by a purely microbial biosphere albeit with seemingly increasing complexity over time. A rich record of this geobiological evolution over most of Earth's history thus provides insights into the remote detectability of microbial life under a variety of planetary conditions. Here we leverage Earth's geobiological record with the aim of (a) illustrating the current state of knowledge and key knowledge gaps about the early Earth as a reference point in exoplanet science research; (b) compiling biotic and abiotic mechanisms that controlled the evolution of the atmosphere over time; and (c) reviewing current constraints on the detectability of Earth's early biosphere with state-of-the-art telescope technology. We highlight that life may have originated on a planet with a different (stagnant lid) tectonic regime and strong hydrothermal activity, and under these conditions, biogenic $CH_4$ gas was perhaps the most detectable atmospheric biosignature. Oxygenic photosynthesis, which is responsible for essentially all $O_2$ gas in the modern atmosphere, appears to have emerged concurrently with the establishment of modern plate tectonics and the emergence of continental crust, but $O_2$ accumulation to modern levels only occurred late in Earth's history, perhaps tied to the rise of land plants. Nutrient limitation in anoxic oceans, promoted by hydrothermal Fe fluxes, may have limited biological productivity and $O_2$ production. $N_2O$ is an alternative biosignature that was perhaps significant on the redox-stratified Proterozoic Earth. We conclude that the detectability of atmospheric biosignatures on Earth was not only dependent on biological evolution but also strongly controlled by the evolving tectonic context.

1. School of Earth & Environmental Sciences,University of St Andrews, St Andrews, Fife, KY16 9TS, United Kingdom

2. Department of Earth, Atmospheric, and Planetary Science, Purdue University, West Lafayette, IN 47906, United States

3. U.S. Geological Survey, Geology, Energy & Minerals Science Center, Reston, VA 20192, United States

4. Department of Geosciences, The Pennsylvania State University, University Park, PA 16802, United States

* ees4@st-andrews.ac.uk

# INTRODUCTION: WHY THE EARLY EARTH?

Earth is currently the only known planet with a biosphere and therefore an important reference point in our search for life on other planets. The ability of Earth's modern biosphere to generate biosignatures that can be detected remotely with a space telescope was famously documented by the Galileo spacecraft in 1990 (Sagan et al. 1993), and this concept continues to underpin the field of Astrobiology to this day. However, it is also widely recognized that the modern biosphere is highly advanced and not necessarily a good analogue for life on other worlds. Earth's atmosphere today contains 21% $O_2$ gas that is generated by cyanobacteria and plants, where the latter contain chloroplasts derived from cyanobacterial ancestors. The origin of cyanobacteria may go back to a singular event in biological evolution (reviewed by Sánchez-Baracaldo and Cardona 2020). Essentially all complex life today depends on $O_2$ gas as an electron acceptor in metabolic energy transformation (see below), but it remains unknown if (a) an $O_2$-producing metabolism such as oxygenic photosynthesis would evolve independently on another planet, and if (b) $O_2$-producing organisms on other planets (if they ever arose) would modify surface environments to a similar extent as they have on Earth, where they enabled the rise of oxygen-breathing macrofauna (Margulis and Lovelock 1974; Margulis et al. 1976). For these reasons it is conceivable, and perhaps likely, that extra-terrestrial life could be entirely microbial and largely anaerobic (Cavicchioli 2002). The search for life on other planets needs to account for this possibility and consider the effects of anoxia on nutrient availability, biological productivity, and biosignature production.

Astrobiologists over the past two decades have therefore begun to use the early Earth, prior to the rise of complex aerobic organisms, as an analogue for inhabited exoplanets (e.g., Des Marais et al. 2002; Benner et al. 2004; Kaltenegger 2017; Catling et al. 2018; Schwieterman et al. 2018; Van Kranendonk et al. 2021). Using geological and geochemical techniques, it is possible to "see through" limited availability and alteration of the ancient rock record to reconstruct environmental conditions and ecosystems in the distant past (further discussed below). For example, phylogeny and knowledge of microbial metabolisms under anoxic conditions based on modern analogues and cell cultures (see below) can provide insights into the types of biosignature gases that were likely dominating Earth's surface environments billions of years ago. These biosignatures need to be evaluated against a backdrop of tectonic and volcanic processes at that time (see below). Importantly, the modern atmosphere is strongly shaped by biological metabolisms, which have significantly sped up the production and recycling of many gaseous species (e.g. Lenton 1998). This strong biological control progressively co-evolved with abiotic planetary processes over the past four billion years. In the following, we will first describe the fundamental principles that guide this line of research. We will show how these geological and biological tools have been applied to build a knowledge base of early Earth and its biosphere. Emphasis will be placed on biological and geological processes that generate remotely detectable gases or that inhibit their production. Finally, we will leverage these records to present quantitative estimates of gas fluxes through time that can place helpful constraints in our search for extra-terrestrial microbial biospheres.

# TOOLS AND PRINCIPLES OF PALEO-BIOGEOCHEMISTRY

## Deciphering biosignatures from the rock record

Reconstructing the history of life and environmental conditions on the early Earth relies on a combination of geological mapping, geochemical analyses, laboratory experiments, and computational models (e.g., Oró et al. 1990). These four methodological pillars are linked to the quality of the rock record, which determines the amount and reliability of information that is preserved over billions of years and ultimately constrains boundary conditions for experiments, models, and our understanding of the evolution of life on Earth. It is therefore crucial to appraise the limitations that the rock record imposes and the tools that are available to read this biogeochemical archive.

There are broadly three categories of limitations that can hinder or prevent accurate assessments of ecosystems and habitats in deep time: diminishing preservation of rocks with increasing age; preservation bias towards a small subset of paleoenvironmental settings; and alteration of geochemical tracers by secondary processes such as metamorphism or infiltration of younger fluids. Approximately 70% of the modern Earth's surface is covered by oceanic crust that is less than 200 million years old and constantly recycled through plate tectonics (Veizer and Mackenzie 2014 and references therein). Hence with the exception of a few rare ophiolites (i.e., slivers of oceanic crust that have been obducted onto continents), we have essentially no record of the abyssal deep ocean that is older than the Mesozoic (Peters and Husson 2017). Older rocks and sediments are only preserved on continents, where they are subject to erosion, weathering and/or amalgamation during tectonic collisions. Due to these large-scale processes, rocks of Precambrian age are exceedingly rare (reviewed by Van Kranendonk et al. 2018). None of them exceed 4 Ga; only a few grains of zircon ($ZrSiO_4$) have survived from the Hadean eon (Harrison 2009), leaving more than 500 million years of Earth's history nearly unsampled. Furthermore, since the onset of modern-style plate tectonics in the late Archean (see below), marine sediments deposited on continental crust (i.e. on the inner or outer shelf) are preserved preferentially, because they are least likely to undergo erosion and subduction (Peters and Husson 2017). Early Archean greenstone successions formed under a different tectonic regime may be an exception to this rule. Many inferences that we can draw about the early biosphere may thus be biased towards processes operating in less than a few hundred meters of water depth.

Another caveat in studying the ancient rock record are the effects of diagenesis, thermal maturation, and metamorphism, which arise from the increase in pressure and temperature in rocks and sediments that occur after initial microbial recycling and lithification as a consequence of continuous sedimentary burial, tectonism, and magmatic heating. Organic matter deposited in sediments first undergoes diagenesis (low-temperature degradation within unmetamorphosed sediments) and ultimately metamorphic heating when it progressively loses functional groups, *i.e.*, it becomes depleted in elements other than carbon (O, H, N, P, etc.). Diagenesis quickly degrades DNA and proteins over the course of hours to months, with a few exceptions of retention for several years, and hence these molecules are not preserved over long geological time scales in sedimentary strata (e.g., Nielsen et al. 2007; Moore et al. 2012; Moore et al. 2014; Parducci et al. 2017; Capo et al. 2022). Following diagenesis, most remaining biological macromolecules are destroyed by metamorphic processes. Lipids are the most recalcitrant components of dead cells and can in some instances be preserved over

million- to billion-year timescales as “molecular fossils” (Love and Zumberge 2021). However, a large fraction of hydrocarbon molecules are typically released when buried sediments reach the “oil window” (70-150 °C) and beyond (Horsfield and Rullkotter 1994). Residual organic matter is converted into kerogen (solid organic matter that is insoluble in non-polar organic solvents) and ultimately, with further increase in pressure, into highly inert graphite (Hayes et al. 1983). Much of the original information content of biomolecules, such as genome sequences or protein content, is therefore missing from Precambrian sedimentary rocks, limiting the geochemical toolbox to techniques that are applicable to kerogenous and graphitic material or morphological features (Gaines et al. 2009). During progressive metamorphism, hydrous minerals such as clays, gypsum or opaline silica tend to lose water and slowly convert into water-poor or anhydrous phases (Stepanov 2021). The expelled fluids may facilitate ion-exchange reactions between solids. Export of such fluids can lead to alteration of bracketing strata, a process called metasomatism (e.g., White et al. 2014).

Morphological evidence in the form of stromatolites, MISS (microbially-induced sedimentary structures) and microfossils are among the most persistent records of life that can survive metamorphism and date back to through most of the sedimentary rock record (e.g., Walter et al. 1980; Allwood et al. 2006; Noffke et al. 2013; Sugitani et al. 2015); however, their information content with regards to specific metabolisms remains limited (further discussed below). Therefore, stable isotopic analysis of major and minor elements contained in ancient biomass and sedimentary minerals is one of the primary tools used in biogeochemical studies that can sometimes shed light on metabolic evolution despite some of the issues noted above. Many of the key elements that life uses and that are cycled through the ocean-atmosphere system have multiple stable isotopes with the same number of protons (and hence identical chemical properties) but different numbers of neutrons. The neutron number changes the mass and thus influences chemical bond strengths and reaction rates in biogeochemical processes. For example, carbon exists as $^{13}C$ and $^{12}C$ (ignoring the cosmogenic, short-lived $^{14}C$ isotope), and the $^{13}C/^{12}C$ ratio of biological organic matter is lower than that of carbonate rocks (Fig. 1), because organisms that perform autotrophic $CO_2$ fixation preferentially uptake the lighter carbon isotope. The fundamental reason for this isotopic fractionation is a slightly faster reaction rate for $^{12}CO_2$ compared to $^{13}CO_2$ as the gas interacts with the enzymatic machinery of the cell (O’Leary 1981). Although this particular isotopic signature of biomass may be altered by metamorphism, as organic matter converts to kerogen and graphite, it is not necessarily lost from the record even up to high metamorphic grade (Hayes et al. 1983; Schidlowski 2001). Complications, however, may arise if the original $CO_2$ is derived from a non-atmospheric source and/or if abiotic processes reduce $CO_2$ to organic matter with similar isotopic effects (see below).

Isotopic data are typically expressed in delta notation, such as $\delta^{13}C$ [‰] = $[(^{13}C/^{12}C)_{sample}/(^{13}C/^{12}C)_{VPDB} - 1] \times 1000$. Here, VPDB stands for Vienne Pee Dee Belemnite, an internationally agreed-upon reference standard that is used for normalizing data from different laboratories to a common denominator. When interpreting isotopic data, an important consideration is isotopic mass balance. As biomass accumulates carbon with a relatively low $\delta^{13}C$ value, due to preferential uptake of $^{12}C$, the residual $CO_2$ in the atmosphere becomes depleted in $^{12}C$ or enriched in $^{13}C$. This signature is inherited by carbonate rocks that form from dissolution of $CO_2$ gas in water (Fig. 1). The more biomass is produced, the stronger the $^{12}C$-depletion in carbonate. Time-resolved analyses of $\delta^{13}C$ in carbonates through Earth’s history therefore hold information about the evolution of biological activity (e.g., Kump and Arthur 1999; Krissansen-Totton et al. 2015; Planavsky et al. 2022). Similarly, stable isotopes of sulfur (preserved in sulfide minerals and in sulfate minerals) can inform us about the evolution of the sulfur cycle through time (Canfield and Farquhar 2009), which has important implications for other nutrients whose solubility is sensitive to sulfur chemistry (see below).

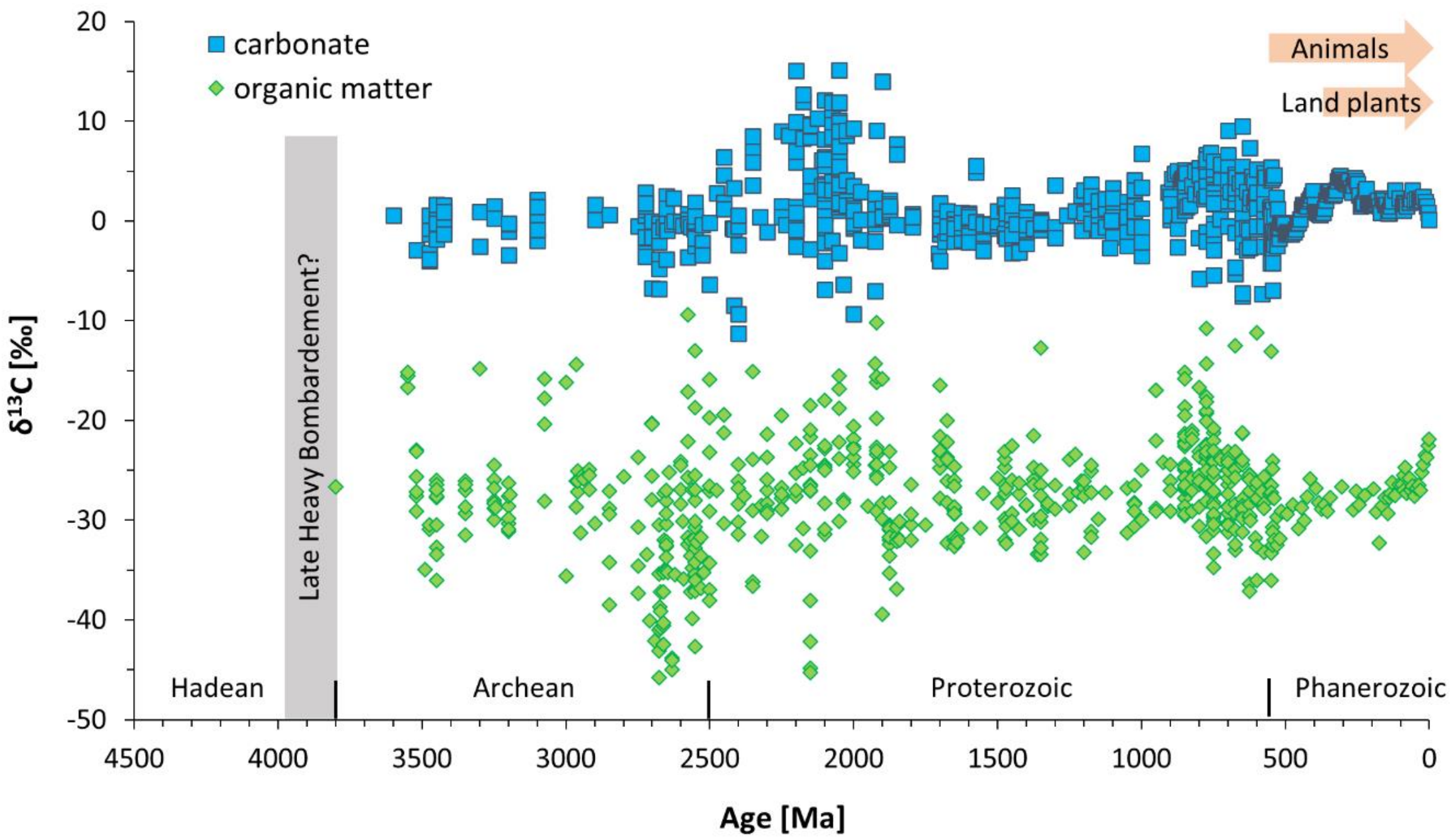

***Figure 1:*** *Inorganic (carbonate) and organic carbon isotopes versus age in millions of years, starting near the formation of Earth. Database taken from Krissansen-Totton et al. (2015), who presented each geological formation by its average composition. The oldest data point at 3.8 Ga is from the Isua Supracrustal Belt (see below) and also marks one of the oldest occurrences of meta-sedimentary rocks on Earth. No rocks are preserved from the Hadean Eon due to tectonic recycling and destruction by impacts.*

In addition to stable isotopes of major and minor elements, the abundances of elements and their speciation in sedimentary rocks can also provide insights into ocean chemistry and nutrient availability. For example, the speciation of iron (*i.e.*, the relative proportions of Fe bound to carbonates, oxides, sulfides, and silicates) has been developed as a means of reconstructing the redox state of ancient water columns locally, above the sediment-water interface of the sampling site (Poulton and Canfield 2005; Raiswell et al. 2019). A more global proxy for the redox state of the deep ocean and marine sediments is the abundance molybdenum in sediments from anoxic settings (Algeo and Lyons 2006; Scott and Lyons 2012; Little et al. 2015). The fundamental premise of this proxy is again based on mass balance. Molybdenum is well-mixed throughout the ocean today, but it is rapidly drawn down into sediments under anoxic conditions, where it reacts with reduced sulfur and becomes particle reactive (Helz et al. 2011). Hence the more expansive anoxia is across the world's oceans, the smaller the Mo reservoir in global average seawater and the lower the concentration of Mo at any given site (Anbar and Knoll, 2002; Scott et al., 2008). It thus becomes possible to make predictions for the entire past ocean, even if only a few outcrops of rocks are preserved from a given time window (Reinhard et al. 2013). Other redox-sensitive elements such as U, As, V, or Cr may be used in similar ways (Tribovillard et al. 2006; Reinhard et al. 2013; Fru et al. 2019). Ultimately, the records of isotopic data, elemental speciation and abundances can be compiled and fed into computational models to derive quantitative estimates of biogeochemical cycles through deep time (see last section). In the following, we will first go through this approach of first summarizing the evidence preserved in the rock record and then converting those observations into quantitative estimates of biosignature gases.

**Fundamentals of biochemistry**

All life on Earth gains energy for growth and metabolism through exothermic electron transfer processes, and this fundamental principle has probably been true since life's origins (Moser et al. 1992). It is therefore conceivable that electrochemistry also forms the basis of extra-terrestrial biospheres. In order to perform electron transfer reactions, organisms must obtain electron donor and electron acceptor substrates from the environment (Moser et al. 1995). The combination of substrates fundamentally differs between chemotrophic and phototrophic organisms. Chemotrophs acquire both electron donors and acceptors from aqueous phases, gases, or minerals. For example, autotrophic methanogens (i.e., chemotrophs) use hydrogen ($H_2$) as the electron donor and $CO_2$ as the electron acceptor to gain energy and produce the waste products methane ($CH_4$) and $H_2O$ (Fig. 2). The net reaction, which incorporates many intermediate biochemical processes, may be written as follows:

$$CO_2 + 4\,H_2 \rightarrow CH_4 + 2\,H_2O + E$$

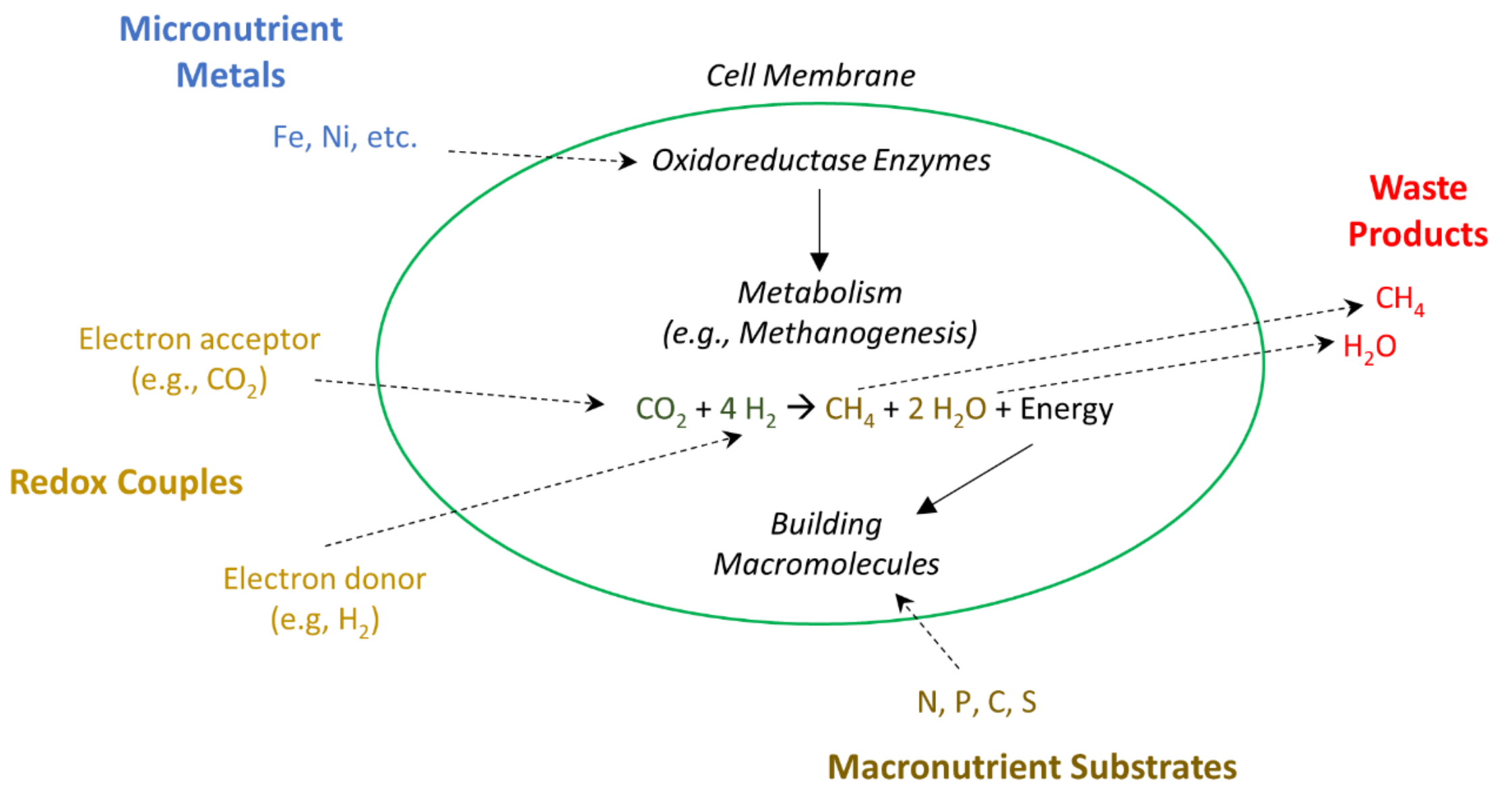

***Figure 2:*** *Schematic of a cell performing methanogenesis with energy and nutrients going in and wastes being excreted.*

The $H_2$ that is used in methanogenesis can come from geological sources, such as hydrothermal vents or from other organisms that produce $H_2$ during anaerobic degradation of organic matter (Kral et al. 1998). The energy gained from this reaction is then used to build complex organic molecules. In contrast, phototrophs use the light energy from the Sun to fix carbon and carry out other metabolic processes (Hohmann-Marriott and Blankenship 2011). The absorption of photons by light-harvesting pigments is optimized to the solar spectrum, meaning that it could differ if a similar light-dependent metabolism were to arise around other stars (Kiang et al. 2007). Oxygenic photosynthesis uses light energy to split water molecules and obtain electrons, which are then used to reduce the carbon in $CO_2$ from the atmosphere and synthesize simple sugars, such as glucose. Here, the net reaction is:

$$6\ CO_2 + 6\ H_2O + \text{Light Energy} \rightarrow C_6H_{12}O_6 + 6\ O_2$$

The simple sugar provides energy and fuels the synthesis of structural polymers such as cellulose, as well as all necessary cellular activities. In aerobic respiration, a particular form of chemotrophy performed by aerobic microbes such as *E. coli* and large animals such as *H. sapiens*, organic matter is the electron donor that is consumed and oxygen ($O_2$) serves as the electron acceptor that is acquired in different ways by different types of organisms (inhalation, absorption, diffusion, etc.). Electrons are transferred from the organic matter (represented by the simple sugar glucose: $C_6H_{12}O_6$) to $O_2$ resulting in the release of energy (E) and the production of carbon dioxide ($CO_2$) and water ($H_2O$):

$$C_6H_{12}O_6 + 6\ O_2 \rightarrow 6\ CO_2 + 6\ H_2O + E$$

This is essentially the same process of burning wood in a fire, where the energy released from a fire is the same type of chemical energy released by the aerobic respiration of organic matter in unicellular or multicellular organisms. The $CO_2$ and $H_2O$ released are the waste products of aerobic respiration. Aerobic respiration generates relatively large amounts of energy and is therefore used by all known macrofauna on Earth (Sperling et al. 2013). In anaerobic respiration, the oxidation of organic matter is coupled to electron acceptors other than $O_2$, such as ferric iron ($Fe^{3+}$), nitrate ($NO_3^-$), or sulfate ($SO_4^{2-}$), which is particularly relevant for anoxic worlds. However, anaerobic respiration yields generally less metabolic energy than aerobic respiration (Schoepp-Cothenet et al. 2012), which may limit the complexity of such organisms.

Besides thermodynamically favorable pairs of electron donors and acceptors, biological activity is also dependent on the availability of other nutrients that are required to build complex cell structures. There are six essential macronutrient elements (known as the CHNOPS elements) that make up 98% of living matter on Earth: carbon (C), hydrogen (H), nitrogen (N), oxygen (O), phosphorus (P), and sulfur (S) (Cockell et al. 2016). Other micronutrient elements are also essential, in trace amounts (e.g., Ni, Mo, Mn, Cu, and Fe). Across the tree of life, oxidoreductase enzymes are the nanomachines that perform electron transfer for obtaining energy and cellular metabolism (Falkowski 2015), and they commonly contain transition metals in their active sites to catalyze electron transfer (Williams 1981).

Transition metal cofactors used in oxidoreductases are required for various metabolic pathways, such as nickel (Ni) for methanogenesis, manganese (Mn) for oxygenic photosynthesis, or copper (Cu) for aerobic respiration. Iron (Fe) is the most widely used metal in biology and is essential for the metabolisms described above, as well as for many others (e.g., sulfate reduction, anoxygenic photosynthesis, nitrogen fixation, etc.). Indeed, large areas of the modern ocean have limited biological growth because of low Fe availability (Behrenfeld et al. 1996; Behrenfeld and Kolber 1999). When essential macronutrient CHNOPS elements or micronutrient metal cofactors are not available in the environment, they limit biological production and the excretion of waste product gases that can be crucial substrates for other organisms (Jelen et al., 2016). The ability of transition metals to catalyze electron-transfer reactions relates to fundamental chemical properties of these elements. Therefore, if life on other planets evolved that also operates on the basis of electrochemistry, where redox reactions are used to gain metabolic energy (Fig. 2), then it is conceivable that transition metals would fulfil a similar catalytic function.

While $O_2$ is essential for aerobic life and thus all macrofauna as we know it, it is simply a waste product of oxygenic photosynthesis. All metabolisms excrete waste products, many of which are gaseous and can leak into the atmosphere: $CO_2$ is a waste product of respiration; hydrogen sulfide ($H_2S$) is a waste product of sulfate reduction; $CH_4$ is a waste product of methanogenesis. When these biological waste product gases accumulate in the atmosphere, they can be identified as planetary biosignatures using remote sensing approaches. For example, the presence of abundant oxygen in Earth's atmosphere is extremely rare in the known Universe surveyed to date (Waltham 2014). Detecting oxygen or other biological waste product gases in the atmospheres of other planets could be a potential sign of life and perhaps a sign that the planet's biosphere has taken control of the residence time of many atmospheric species to a similar extent as it has on Earth. Understanding past and present planetary constraints on the biological production of these waste products (above non-biological background emissions) may therefore be crucial for detecting alien biospheres.

## Fundamentals of geological constraints on Earth's atmosphere

As outlined above, biological processes can have a significant, and potentially observable, effect on a rocky planet's atmosphere. On the modern Earth, the atmospheric composition is very strongly controlled by life (Lenton 1998). However, any potential atmospheric biosignature must be disentangled from a backdrop of abiotic (geological and astrophysical) processes that also contribute to planetary atmospheres and would be dominating on lifeless worlds and on planets with a very small biosphere. Planet formation models (Ikoma and Genda 2006; Lee et al. 2014) and observations of exoplanet populations (Fulton et al. 2017) suggest that nascent planetary embryos accrete primary atmospheres from proto-planetary nebula gas, predominantly hydrogen and helium. While it is not conclusively known if Earth ever possessed such a primary atmosphere, there is evidence from noble gases of ingassing of nebula gases during formation (Williams and Mukhopadhyay 2019; Lammer et al. 2020). Moreover, Young et al. (2023b) argue that the light element abundance in Earth's core, origin of Earth's water, and the oxidation state of the mantle can all be explained by the interaction of the proto-Earth with an $H_2$-He atmosphere. However, each of these aspects of the Earth are also well explained without primordial atmosphere ingassing (see reviews by Frost and McCammon 2008; Li and Fei 2014; Peslier et al. 2017), so should not be taken as direct evidence of an $H_2$-He atmosphere for the proto-Earth. The purported primary atmosphere must have been lost soon after planet formation. Proposed mechanisms for primary atmosphere loss include photoevaporation from the host star (Owen and Wu 2013), atmospheric blow off driven by the interior heat of the planet (core powered mass loss, Ginzburg et al. 2018), or impact stripping (Genda and Abe 2003; Schlichting et al. 2015).

Earth's so-called secondary atmosphere (the precursor to our current atmosphere) then formed from outgassing of volatiles stored in the interior (Schaefer and Fegley 2017; Gaillard et al. 2021). These would primarily have included different species of carbon ($CO_2$, CO, or $CH_4$), hydrogen ($H_2$, $H_2O$), nitrogen ($N_2$, $NH_3$), and sulphur ($S_2$, $SO_2$, $H_2S$) gases, with redox state depending on the oxidation state of Earth's mantle. Earth's mantle is highly oxidized today, and thus $CO_2$, $N_2$, $H_2O$ and $SO_2$ dominate over CO, $CH_4$, $NH_3$, $H_2$ and $H_2S$ (Schaefer and Fegley 2017; Gaillard et al. 2021). Geochemical evidence indicates that Earth's mantle has been oxidized since the Archean (Frost and McCammon 2008), and possibly even earlier (Trail et al. 2011). How Earth developed this oxidation state is not fully understood, but leading theories point to disproportionation of iron oxide at high pressure during the magma ocean phase, where a significant fraction of the mantle was molten due to heat from accretion, short-lived radioactive isotopes, or giant impacts (Wade and Wood 2005). Iron metal formed by disproportionation then rained down to join the core, leaving more oxidized material behind in the mantle (Wade and Wood 2005). The disproportionation mechanism applies to planets Earth-size or larger, as these planets reach high enough pressures in their lower mantles for disproportionation to occur. If this mechanism is correct, then we might expect super-Earth exoplanets to also have oxidized mantles and outgas relatively oxidized volatile species.

Significant degassing of $N_2$, $CO_2$, $H_2O$ and $SO_2$ likely occurred during the early magma ocean state, in the wake of the moon forming impact, giving rise to Earth's secondary atmosphere (Sleep et al. 2014). The composition and size of the atmosphere were at that time, prior to the emergence of life, controlled by thermodynamic equilibrium between the atmosphere and the magma ocean (Elkins-Tanton 2008; Gaillard et al. 2022; Wolf et al. 2023). After the magma ocean solidified and Earth progressed towards a habitable state where $H_2O$ was condensed as liquid water on the surface, volcanism became the main abiotic source of volatiles to the atmosphere. Volcanism has persisted on Earth throughout its history, which may have included tectonic regimes different than modern day plate tectonics, including a stagnant-lid regime similar to that which occurs on Mars (Korenaga 2013; O'Neill and Debaille 2014; Stern et al. 2018). The long-term equilibrium state of the atmosphere is controlled by volcanic and ultimately biological sources of atmospheric gases and sinks through processes such as atmospheric escape (applies to light species like H), biological uptake, or interaction with crustal rocks as occurs during weathering or hydrothermal alteration (e.g., Catling 2014).

Interaction between the atmosphere and the crust generates new volatile-rich minerals such as carbonates, hydrated silicates, ammonium-bearing clays, and sulfide or sulfate deposits. The biosphere likely enhanced some of these processes once it emerged, for example by promoting carbonate precipitation or sulfide deposition (see also Hazen et al. 2008) and by storing volatiles in the form of organic molecules. During plate tectonics, as on the modern Earth, these volatile-rich phases can be returned to the mantle by subduction or they are re-melted to drive further degassing (Bekaert et al. 2021). Tectonics thus recycles volatiles between the surface and interior, thereby providing a continual source of volatiles to the mantle to allow degassing to persist as long as volcanism is active. Life indirectly contributes to this flux.

Volcanism on the modern Earth occurs primarily at mid-ocean ridges, at hot spots, and at subduction zones, and these are likely the dominant volcanic settings on any rocky planet in a plate-tectonic regime. Volcanism will also occur on planets in a stagnant-lid regime, given sufficient mantle heat sources, as discussed below. Volcanism at mid-ocean ridges occurs when upwelling mantle, rising adiabatically, crosses the solidus and melts. This source of volcanism persists as long as the mantle temperature is higher than the mantle solidus at surface pressure. Hot spots are thought to be typically caused by upwelling mantle plumes, which are hotter than the surrounding mantle and drive volcanism at plate interiors where plumes impinge upon the base of the lithosphere. Volcanism at subduction zones occurs when down-going plates sink into the mantle and heat up to the point where the sinking crust either melts itself, or releases water that has been incorporated in the subducting plate and associated sediments by hydrothermal circulation to the overlying mantle, thereby lowering the solidus and inducing melting. Tectonics and volcanism on Earth are thus important factors in regulating habitability and atmospheric evolution over many hundreds of millions of years, i.e. on timescales longer than those over which biological metabolism operates and internally recycles volatiles.

## THE GEOLOGICAL RECORD OF LIFE AND ENVIRONMENTS

### Life on a fully anoxic planet (ca. 4.5-3 Ga)

***Early tectonic constraints.*** Tectonics on the very early Earth (4.5-3.0 Ga, Condie 2007) is poorly constrained due to a sparse rock record that is difficult to uniquely interpret in terms of tectonic processes. The Jack Hills zircons, the only solid remnants from prior to ~4.0 Ga, indicate that liquid water oceans formed during the Hadean based on the oxygen isotope composition of the zircons (Mojzsis et al. 2001; Valley et al. 2002). The Jack Hills zircons were derived from felsic (silica-rich) crust, now lost to weathering. However, other regions of felsic crust are preserved from the very early Earth, including the oldest intact whole rocks on Earth, the ~4.0 Ga Acasta gneiss complex, and represent the first building blocks of the continents (Bowring et al. 1989; Reimink et al. 2019). These preserved regions of continental crust only demonstrate the minimum amount that must have existed at this time; the exact amount of continental crust present through the Hadean and Archean can only be estimated from geochemical models and is debated (e.g., see review by Harrison 2009, in particular their Figure 1). Some authors argue for rapid growth of continents during the early Archean or even Hadean (Korenaga 2018), while others argue for a more gradual growth of continental crust during the Archean, reaching ~75% the present day volume by the end of the Archean (Harrison 2009). The early crust that formed at this time may have been largely submerged, as predicted from isostatic models (Flament et al. 2008; Lee et al. 2018). That continents on the very early Earth were mostly submerged is supported by shale-forming basins not appearing in the geologic record until the end of the Archean (Altermann and Nelson 1998; Knoll and Beukes 2009; Liebmann et al. 2022), and analysis of zircon age distributions arguing for continental emergence above sea level at ~3.0 Ga (Reimink et al. 2021). Hence the earliest life forms likely emerged and evolved on an ocean-dominated world, where the only land masses were perhaps volcanic islands above hotspots, akin to Hawai'i on the modern Earth (Van Kranendonk et al. 2015; Bada and Korenaga 2018). This carries implications for exoplanet science, because it means that an independent origin of life may not necessarily require the presence of large land masses on the scale what we experience today.

Formation of the continental crust as seen on the early Earth is thought to require melting of hydrated mafic crust (like ocean crust) at pressures of ~1-3 GPa (~30-100 km depth) (e.g., Moyen and Martin 2012). Therefore, some form of crustal recycling must have been active. However, it is not clear if this was crustal burial by volcanism, as can occur in a stagnant-lid regime (Bédard 2006; Willbold et al. 2009; Reimink et al. 2014), or by subduction (Komiya et al. 1999; Nutman et al. 2002; Moyen and Van Hunen 2012; Martin et al. 2014; Turner et al. 2014). The dominant crust generation process may have switched from a stagnant-lid to a mobile-lid, with subduction, process during the Archean (Dhuime et al. 2012; Bauer et al. 2020), or even varied spatially across the Earth (Van Kranendonk 2010). As stagnant-lid tectonics is a possibility for the early Earth, we consider here how volcanism and volatile cycling would operate in such a regime. If the early Earth was in a stagnant-lid regime, the lithosphere would have been significantly thicker than in a plate-tectonic regime, forcing melting to occur at a greater depth. The mantle must be hotter in order to melt at a greater depth, as the solidus temperature increases with pressure, meaning that volcanism is impeded on stagnant-lid planets (Solomatov and Moresi 1996; Reese et al. 1999). However, stagnant-lid planets also lose heat less efficiently than plate-tectonic planets, meaning they tend to have hotter mantle temperatures. As a result, volcanism is expected to still be active on such planets given sufficient heat sources (Foley et al. 2020; Unterborn et al. 2022), and as evidenced by recent volcanism on Mars (Hartmann et al. 1999; Hauber et al. 2011), meaning volatile outgassing occurs as well. Venus also shows evidence for recent volcanism (Smrekar et al. 2010; Shalygin et al. 2015), and while it does not possess plate tectonics like the modern Earth, it is probably not in a true stagnant-lid state either (Davaille et al. 2017; Borrelli et al. 2021; Smrekar et al. 2023). Ultimately, even if the very early Earth lacked plate tectonics, there would still have been significant abiotic fluxes of $CO_2$, $N_2$, $H_2O$ and $SO_2$. Substantial mantle melting and volcanism on the very early Earth is also supported by the geologic record. For example, the presence of komatiites is thought to represent very high degrees of mantle melting (Viljoen and Viljoen 1969; Brooks and Hart 1974), which is also supported by models of melt production through time based on global databases (Keller and Schoene 2012; Keller and Schoene 2018).

In theory, a planet that operates in a stagnant-lid regime and hence lacks subduction, as may have been the case for the early Earth, also loses the primary mechanism for recycling of volatiles from the surface back to the interior. This means that outgassing may be a “one-way street” in a stagnant-lid regime, where volatiles are outgassed to the surface initially and until these volatiles become depleted in the mantle. However, one possible mechanism for recycling surface volatiles to the interior in a stagnant-lid regime is burial of surface rocks by progressive eruption, which, if subaqueous, would take the form of pillow lavas. These eruptive lavas will be extensively altered and weathered by interaction with surface water, hydrating and carbonating them (Foley and Smye 2018; Marien et al. 2023). If surface rocks are buried deep enough, they can sink into the mantle, returning any volatiles acquired by weathering or hydrothermal alteration (Elkins-Tanton et al. 2007). During this burial process, though, rocks will experience increasing temperatures and pressures, and metamorphic reactions may devolatilize rock before it can founder back into the mantle (Foley and Smye 2018). Hence volatiles may have only been able to recycle between the surface and crust if the early Earth was in a stagnant-lid regime, though outgassing of volatiles to the atmosphere would be active regardless.

Another feature of the hot, volcanically active, early Earth with significant volumes of (ultra-)mafic lavas would have been the likely presence of intensive hydrothermal activity on the seafloor, as suggested by trace element data (Viehmann et al. 2015). Hydrothermal vents are considered important sites for the origin of life, because they provide catalytic mineral surfaces and reduction potential through the process called serpentinization (Baross and Hoffman 1985; Martin et al. 2008; Russell et al. 2010). At these sites, $H_2O$ reacts with $Fe^{2+}$ present in ferrous minerals such as olivine and pyroxene, forming $H_2$ gas. In the presence of magnetite or iron sulfides, $H_2$ can reduce $CO_2$ to $CH_4$ and longer-chain hydrocarbons (McCollom and Seewald 2007; Proskurowski et al. 2008), and $N_2$ can be reduced to $NH_3$ (Brandes et al. 1998). Hence volcanism may have acted as an important engine of prebiotic networks. The generally reducing conditions in these settings create a redox gradient relative to the more oxidized gases released from the mantle, which may have been beneficial for driving prebiotic reactions (Martin et al. 2008). The mantle oxidation state implies that oxidized gases would have been released by volcanoes, meaning reducing gases required a different source. Serpentinizing hydrothermal vents is one possibility; another possibility is outgassing of impactors (Zahnle et al. 2020). Planetary building blocks in the inner solar system have a lower oxidation state than the modern Earth's mantle, so degassing of these materials upon impact can produce strongly reducing atmospheres that could have persisted for ~1-10 Myrs (Zahnle et al. 2020).

***Traces of early life in the Archean (up to ca. 3 Ga).*** This framework of a volcanically active planet with limited continental exposure provides context for the oldest vestiges of life on Earth. The oldest putative signs of life in the rock record, though controversial, date back 4.1 Ga and are preserved in the form of graphite inclusions in a single zircon crystal (Bell et al. 2015). Zircons crystallize from magma in the subsurface, and so the interpretation of the carbon in the graphite inclusion is that it was originally deposited as biomass within marine sediments, which were subsequently buried to several kilometres depth where the biomass was converted to graphite as a consequence of heating and compression. During further burial and heating, the sediment package ultimately underwent melting and zircons formed that trapped some of this graphite as inclusions. The entire melt ultimately solidified into a magmatic rock, which was later exhumed and eroded, releasing the zircon grain back into the environment where it was found. The key argument for the biogenicity of the graphite inclusions within this zircon is its carbon isotope composition ($\delta^{13}C$) of -27‰, which is indistinguishable from average modern biomass (Schidlowski 2001). This carbon isotope value could possibly reflect an autotrophic lifestyle with biological $CO_2$ fixation, but also a heterotrophic metabolism, where living organisms were feeding on abiotically produced organic matter, cannot be ruled out from these data alone. However, the similarity to modern autotrophic organisms leaves open the possibility that autotrophy was active by 4.1 Ga. If so, it would be the oldest evidence of life impacting the composition of Earth's atmosphere, because autotrophic $CO_2$ fixation would create an additional flux of $CO_2$ from the atmosphere into the lithosphere as organic matter underwent burial.

However, this conclusion remains controversial due to the small size and singular occurrence of Hadean zircon-hosted graphite (Nemchin et al. 2008). Also the carbon isotope value could theoretically be mimicked by abiotic processes (McCollom and Seewald 2006). If correct, the presence of life in the Hadean at 4.1 Ga or earlier would have important implications for exoplanet science, because it would show that life was able to emerge on Earth during the period of relatively heavy meteorite bombardment (Bottke and Norman 2017). Although it has been called into question if Earth experienced an episode of increased impact rates ("Late Heavy Bombardment") at around 4.0-3.8 Ga (Boehnke and Harrison 2016), impact rates were still likely higher in the early history of the solar system (Bottke and Norman 2017; Lowe and Byerly 2018), and the same would likely be true for other solar systems with rocky planets. If life was able to originate on Earth despite frequent impacts at that time, it would thus bode well for the habitability of terrestrial planets elsewhere. As noted above, meteorite bombardment may create a linkage between the origins of life and the reducing power imported by impact events (Zahnle et al. 2020). The development of life at this time may also imply that the biosphere had enough refugia available to survive impact events, for example in the subsurface (Abramov and Mojzsis 2009), and/or it could mean that life went extinct and re-emerged multiple times, meaning that biogenesis as such is perhaps not rare.

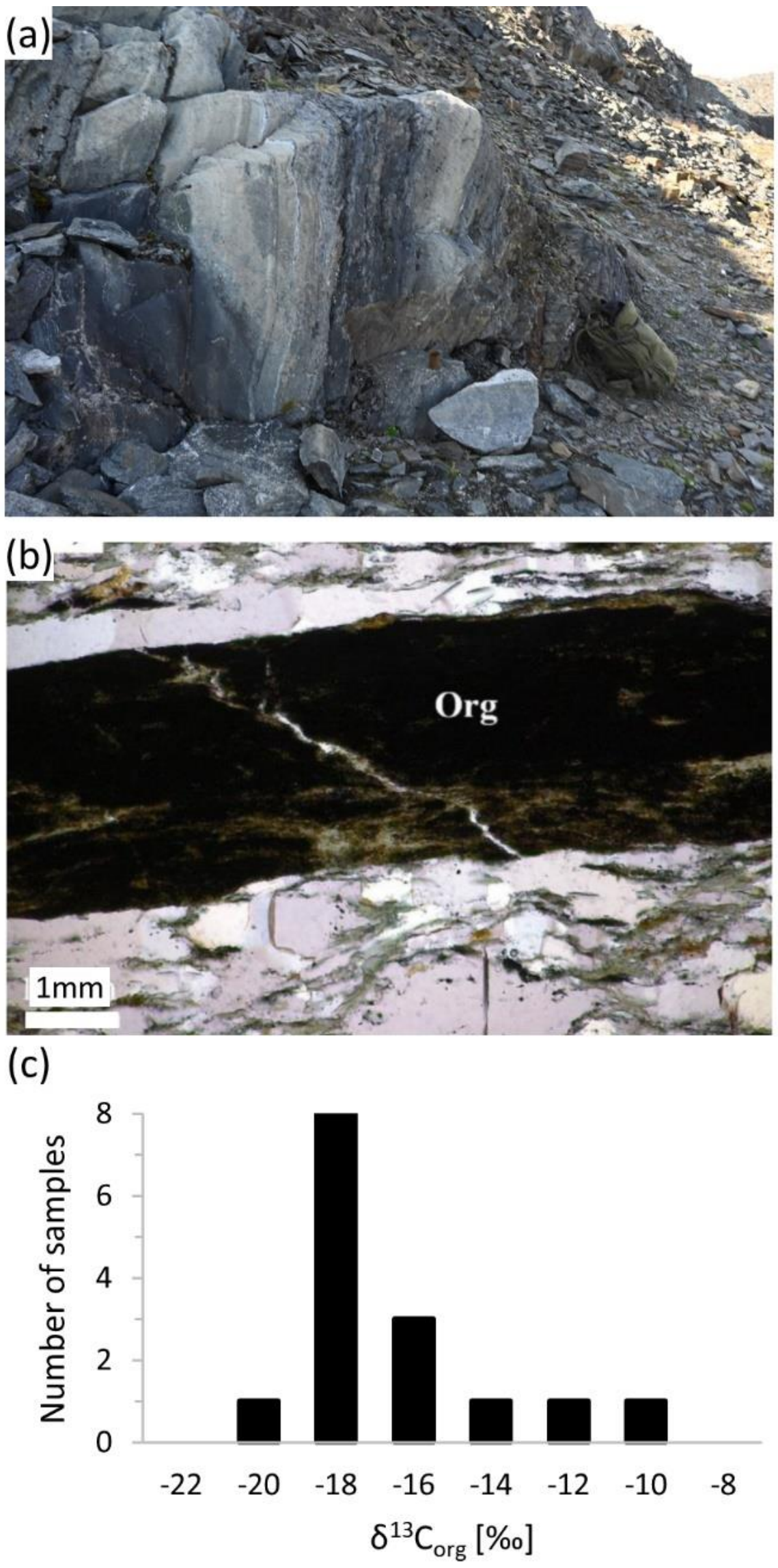

***Figure 3:*** *Organic carbon isotopes in graphitic schists as a potential biosignature. Shown here as an example is the Garbenschiefer Formation from the Isua Supracrustal Belt in Western Greenland. (a) Outcrop photo of the meta-turbidite (reproduced from Stüeken et al. 2021a, under Creative Commons Attribution License CC-BY). Backpack on the right side of the outcrop is approximately 50 cm tall. (b) Photomicrograph showing graphite lamina in quartz-mica schist (image credit: Jane Macdonald, St Andrews, 2023). (c) Organic carbon isotope data from these rocks, replotted from Rosing (1999) and Stüeken et al. (2021a).*

The next oldest and less controversial evidence of life on Earth comes from graphite-rich laminae within highly metamorphosed sedimentary rocks in several localities in the Isua supracrustal belt in Greenland, dated to approximately 3.7-3.8 Ga (Mojzsis et al. 1996; Rosing 1999; Ohtomo et al. 2014; Hassenkam et al. 2017) (Fig. 3). Although slightly less negative than the graphite inclusions in zircon, possibly due to metamorphic overprinting, this metasedimentary graphite shows $\delta^{13}C$ values that are consistent with biological $CO_2$ fixation. Of similar age are putative microfossils preserved in jasper in the Nuvvuagittuq supracrustal belt in Quebec, Canada (Dodd et al. 2017; Papineau et al. 2022) and occurrences of stromatolites in carbonates at Isua in Greenland (Nutman et al. 2016), although the biogenicity of both of these latter occurrences has been called into question (Allwood et al. 2018; McMahon 2019; Lan et al. 2022).

The Eoarchean graphite from Isua and surrounding minerals are moderately enriched in nitrogen (Pinti et al. 2001; Papineau et al. 2005; Stüeken 2016; Hassenkam et al. 2017). These N-enrichments are further support for a biogenic origin of this graphite, because all biomass on Earth contains nitrogen in the form of amine groups as a key constituent of many essential biomolecules, and this nitrogen gets released as ammonium into sedimentary pore waters after deposition on the seafloor and is subsequently trapped in clays (Müller 1977; Schroeder and McLain 1998; Barth et al. in prep.). Hence clays can carry nitrogen originally derived from biogenic organic matter. Biomass burial therefore constitutes a major flux of nitrogen from surface environments into the rock record and affects the pressure of $N_2$ gas in the atmosphere over geological timescales (Som et al. 2012; Som et al. 2016; Johnson and Goldblatt 2018). Today, the conversion of $N_2$ to bioavailable ammonium is largely carried out by microorganisms; merely a few percent of $N_2$ fixation is abiotic, driven by lightning reactions in the atmosphere and by hydrothermal processes (Brandes et al. 1998; Navarro-González et al. 1998). During the origin of life, these abiotic sources of fixed N would have been dominant, and they may have involved gaseous products such as HCN, NO and $N_2O$ that are associated with life on the modern Earth and that may leave signs (and potential false positives) in atmospheric spectra (Barth et al. in prep.). However, at some point during early evolution, life appears to have taken control over its fixed N supply by inventing nitrogenase enzymes that convert $N_2$ into bioavailable $NH_4^+$. It is conceivable that biological $N_2$ fixation (termed diazotrophy) is as old as the Last Universal Common Ancestor (LUCA) of all life on Earth and was thus potentially active at 3.8 Ga (Weiss et al. 2016; Garcia et al. 2020; Parsons et al. 2020) or even 4.1 Ga (Bell et al. 2015); however, the geochemical data from this time period are too heavily metamorphosed to test this proposition. The first geochemical evidence for diazotrophs is from rocks of low metamorphic grade at 3.2 Ga (Stüeken et al. 2015; Homann et al. 2018) and at 3 Ga and younger (Koehler et al. 2019; Ossa Ossa et al. 2019).

Although multiple lines of evidence hint at the presence of life at 3.7-3.8 Ga, the high metamorphic grade of these rocks leaves residual doubt on those results. The most widely accepted evidence of life on Earth therefore dates back to approximately 3.5 Ga (Walter et al. 1980; Buick 2007a; Baumgartner et al. 2019; Lepot 2020; Westall et al. 2023) and is based on a range of proxies, including carbon isotopes ($\delta^{13}C$), sulfur isotopes ($\delta^{34}S$) and stromatolites (Fig. 4a). The sulfur isotope data represent the oldest evidence of biological sulfate reduction, using aqueous $SO_4^{2-}$ that was generated by photochemical conversion of volcanogenic $SO_2$ gas (Ueno et al. 2008; Baumgartner et al. 2020a). These data are thus evidence of volcanic and photochemical processes providing electrochemical energy for the biosphere. Chemotrophic organisms would have been able to harness this energy to generate biomass and proliferate.

Perhaps most important for exoplanet science is isotopic evidence in support of biological methane production (methanogenesis) in rocks at 3.5 Ga (Ueno et al. 2006). This study found that methane was preserved in the form of gas inclusions within quartz crystals that are thought to have precipitated on the seafloor, and that it carried an isotopic fingerprint of biological formation. The stromatolites may reflect the presence of phototrophic organisms (see below). The early Archean biosphere at 3.5 Ga thus appears to already have been relatively diverse (Van Kranendonk et al. 2021). In only slightly younger rocks from 3.5-3.4 Ga, carbonaceous microfossils are preserved (Fig. 4b), providing the first physical evidence of what life looked like at that time (Westall et al. 2001; Westall et al. 2006; Wacey et al. 2010; Sugitani et al. 2015; Flannery et al. 2018). Furthermore, individual microfossils from ca. 3.4 Ga carry isotopic signatures indicative of methanogenesis (Flannery et al. 2018; Schopf et al. 2018). Like microbial sulfate reduction to sulfide, methanogenesis is a chemotrophic metabolism, where either $CO_2$ is reduced to $CH_4$ (chemoautotrophy) or $CH_4$ is liberated during the degradation of organic matter (chemoheterotrophy). In any case, life was capable of generating a greenhouse gas and potential atmospheric biosignature by at least 3.4 Ga.

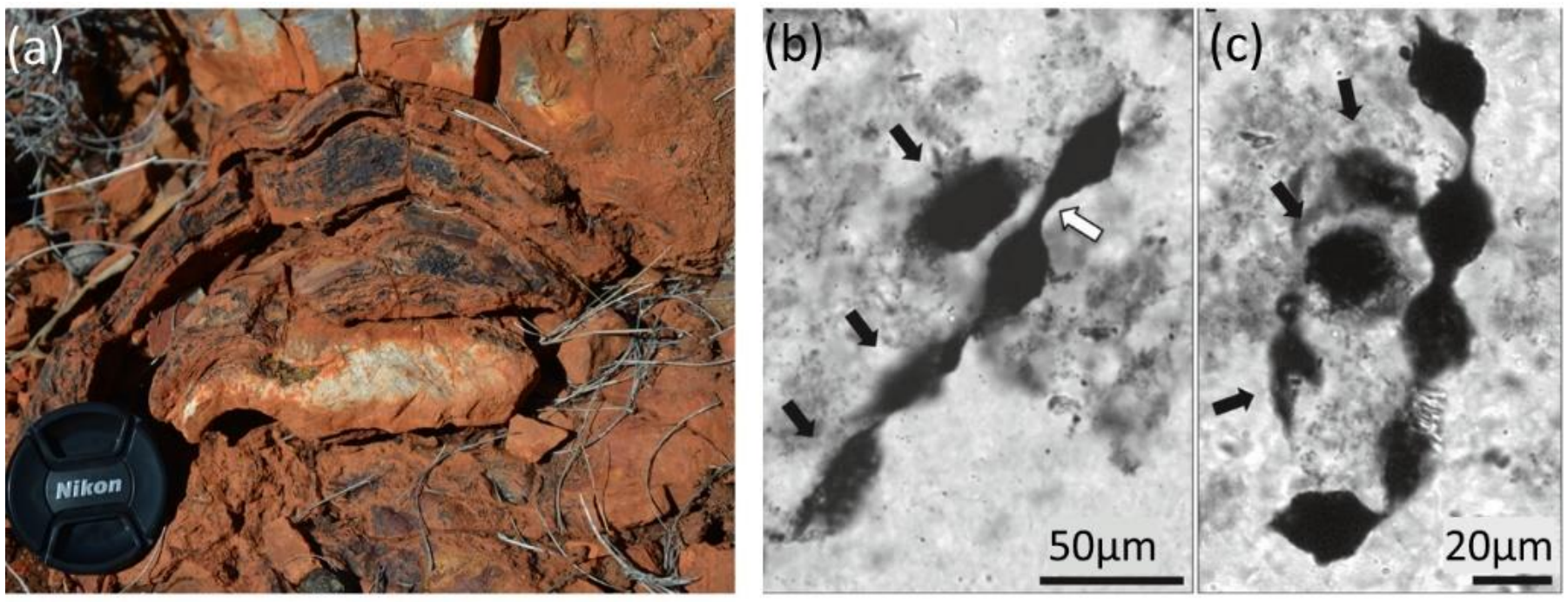

***Figure 4:*** *(a) Stromatolite from the Dresser Formation at 3.49 Ga (photo taken by Eva E. Stüeken). (b) and (c) Photomicrograph of carbonaceous microfossils from the Strelley Pool Formation at 3.4 Ga, reproduced from Sugitani et al. (2015) with editorial permission. White arrows point at a flange-connected pair of cells; black arrows point at isolated lenses.*

Modelling suggests that the concentration of methane in the early Archean atmosphere was likely orders of magnitude above modern values (Catling et al. 2001) (see also below). The reasons for this are increased biological sources paired with a long methane lifetime from very low levels of $O_2$, which allowed methane to accumulate in the atmosphere. Evidence for low atmospheric $O_2$ levels comes from the preservation of mass-independent sulfur isotope fractionation (S-MIF) in the Archean sedimentary rock record (Farquhar et al. 2000; Johnston 2011) (Fig. 5). This signature has been demonstrated to be produced by UV-photolysis of volcanogenic $SO_2$ gas, which requires the absence of a UV-shielding ozone layer. The products of the reaction are $SO_4^{2-}$ (dissolved in rainwater) and elemental S. The preservation of the latter, in particular, requires low levels of $O_2$ in rainwater and seawater, such that oxidation of $S^0$ and homogenization of all sulfur as $SO_4^{2-}$ are prevented. Computational models of this process indicate that the formation and preservation of S-MIF places an upper bound on $pO_2$ of $10^{-5}$ times present atmospheric levels (PAL) (Pavlov and Kasting 2002; Kurzweil et al. 2013).

***Ocean chemistry and nutrient limitation on the Archean Earth.*** As summarized above, from 3.8 to 3.4 billion years ago, geochemical evidence exists for the presence of various types of chemotrophy and phototrophy, which would have had specific metal cofactor requirements (Moore et al. 2017, and references therein). The rate at which biosignature gases were released into the atmosphere would therefore have been limited by the supply and solubility of nutrients and redox couples in the environment at that time. Sedimentary records of iron speciation indicate that the early ocean was anoxic and ferruginous ($Fe^{2+}$-rich), and this state likely persisted for most of Earth's history (*i.e.*, until long after the evolution of oxygenic photosynthesis in the early-/late Archean, see below) (Planavsky et al. 2011; Poulton and Canfield 2011). Under widespread ferruginous marine conditions that prevailed throughout much of Earth's history, photoferrotrophy would have likely been an important form of biological primary production (Canfield et al. 2006; Kendall et al. 2012). Indeed, abundant photoferrotrophs represent a major component of the phototrophic community in ancient ferruginous ocean analogue environments (Crowe et al. 2008; Walter et al. 2014; Llirós et al. 2015). One of the main processes that has been proposed for the origin of banded iron formations (BIFs, chemical precipitates of alternating $SiO_2$ and iron minerals from seawater) prior to the GOE is direct Fe-oxidation by anoxygenic phototrophs (Konhauser et al. 2002; Kappler and Newman 2004). A global redox model has indicated that the combination of $Fe^{2+}$-based and $H_2$-based anoxygenic photoautotrophy in the environment results in amplification of Earth's methane cycle and major influence on climate (Ozaki et al. 2018).

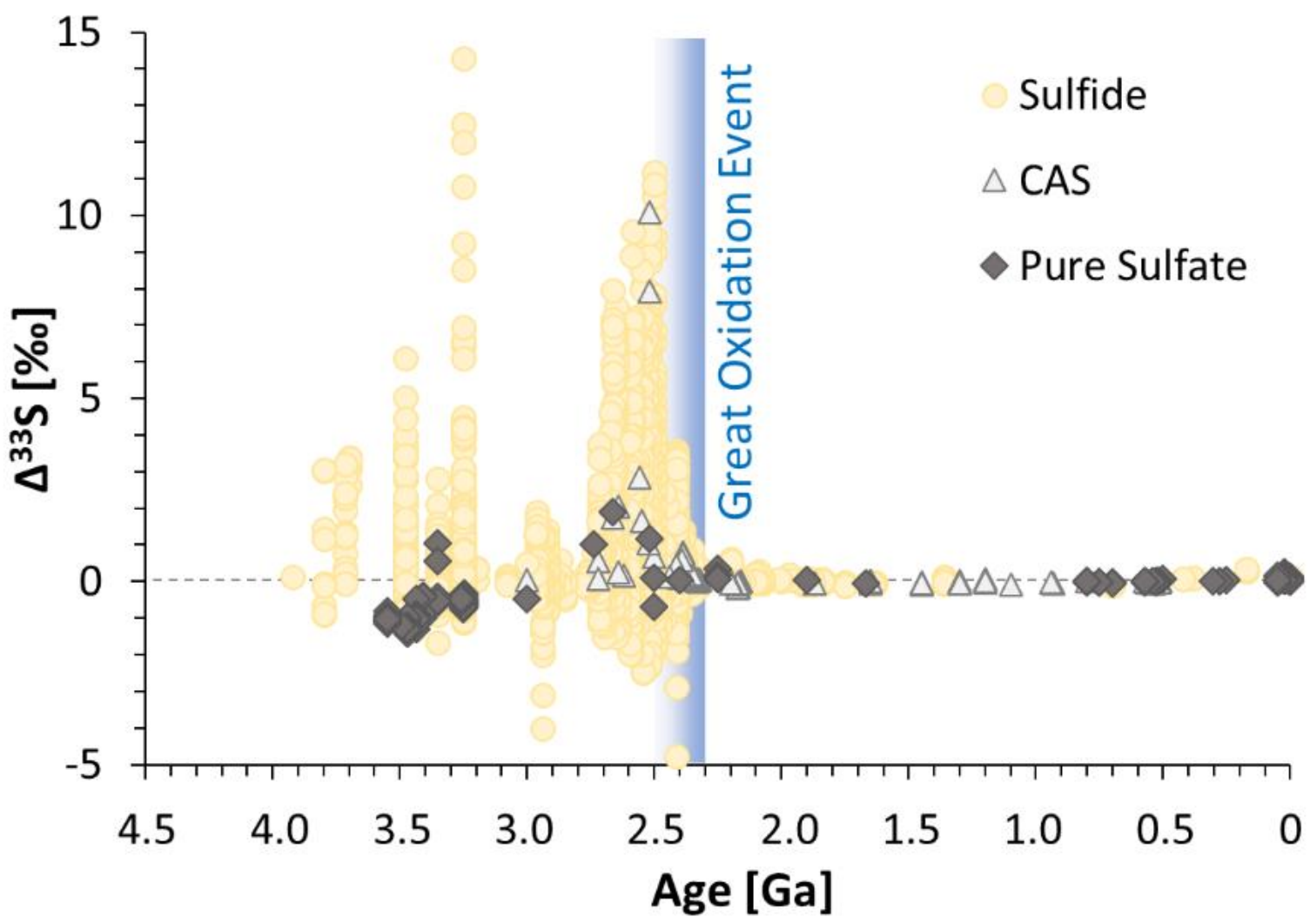

***Figure 5:*** *The sedimentary record of mass-independent sulfur isotope fractionation. Database taken from Claire et al. (2014). The disappearance of the $\Delta^{33}S$ signature at around 2.4-2.3 Ga marks the oxygenation of the atmosphere, which led to the formation of an ozone shield that blocked UV radiation and stopped $SO_2$ photolysis (Farquhar et al. 2000). This event is known as the Great Oxidation Event (GOE). Anoxic planets where $O_2$ production never occurs are more likely to resemble the early Earth prior to the GOE.*

Anaerobic microbial metabolic pathways in the early ocean would have faced different availabilities of essential macro- and micronutrients compared to today (Anbar and Knoll 2002; Saito et al. 2003; Zerkle et al. 2005; Anbar 2008; Robbins et al. 2016). The micronutrient metals Fe, Mn, Ni and cobalt (Co) were present in higher concentrations, while Mo, Cu, and zinc (Zn) were present in lower concentrations (Fig. 6). Iron, Mn, Co and Ni (Rudnick and Gao 2014; White and Klein 2014; Ptáček et al. 2020) are relatively more abundant in mafic crustal rocks, and their solubilities are increased in the absence of oxygen (Fe, Mn) or not strongly affected by redox conditions (Co, Ni) (Brookins 1988). This explains their elevated marine concentrations in the early Archean when the crust was overall more mafic and leached by strong hydrothermal activity (Isley and Abbott 1999; Viehmann et al. 2015). In contrast, Mo, Cu and Zn were likely scavenged from the water column by small amounts of sulfide produced through biotic or abiotic pathways. Only a significant supply of organic ligands may have overcome extreme limitation of Cu and Zn, in particular (Robbins et al. 2016; Stüeken 2020). These trends likely impacted the biogeochemical cycles of other elements and therefore total biological activity and biosignature production in the early Archean (see also Jelen et al. 2016).

Regarding the major nutrient phosphorus (P), the available data are currently inconclusive. Compilations of P abundances in marine shales through time suggest lower marine P availability throughout the Precambrian until ca. 0.7 Ga (Reinhard et al. 2017b; Planavsky et al. 2023). Furthermore, P recycling from biomass would likely have been suppressed under anoxic conditions (Kipp 2022). If so, then P may have been a limiting nutrient over a large geographic region or in individual basins. In contrast, recent experiments have shown that phosphate is moderately soluble in the presence of dissolved $Fe^{2+}$, and therefore phosphate may have been sufficiently bioavailable to support emergent cellular systems and early microbial life (Brady et al. 2022). This interpretation is supported by elevated P levels in marine shelf-ramp carbonates of Neoarchean age, implying sufficient P availability in shallow marine habitats at that time (Ingalls et al. 2022). Similarly, thermodynamic and kinetic modelling indicates that Saturn's moon Enceladus has habitable phosphorus concentrations close to modern Earth seawater (Hao et al. 2022). Hence it is possible that P is not always the critical constraint for the production and detectability of biosignatures, although it is the limiting ingredient for biospheric productivity in most environments today.

Regarding nitrogen, the second major nutrient, redox conditions in the primitive ocean likely directly impacted nitrogen fixation and the nitrogen cycle. In particular, reduced nitrogen is a crucial component of amino acids, proteins, DNA, and RNA. Today, nitrogen makes up nearly 80% of the atmosphere in the form of dinitrogen ($N_2$) gas, but the two nitrogen atoms in dinitrogen gas are connected by a very strong triple bond, making dinitrogen unusable to most organisms. In the absence of significant abiotic sources, bioavailable nitrogen may have been limiting to the earliest microbial communities (Canfield et al. 2010). One may thus speculate that the early- to mid-Archean invention of the nitrogenase enzyme that can break the dinitrogen triple bond and convert $N_2$ into ammonium ($NH_4^+$) (see above) possibly represented a major turning point in the expansion of the biosphere. This hypothesis remains to be tested. Nitrogenase requires several Fe atoms in its structure (Georgiadis et al. 1992), and the most common and most efficient form of nitrogenase includes Mo as an essential metal cofactor along with Fe (Joerger et al. 1988; Garcia et al. 2022). Although Mo can be replaced by vanadium (V) or Fe, phylogenetic data suggest that the Mo-variety of nitrogenase is the most ancient (Parsons et al. 2020). In today's oxic ocean, Fe is the limiting element in the nitrogen cycle (Canfield et al. 2010). However, in the Archean, nitrogen fixation may have been limited by Mo scarcity (Anbar and Knoll 2002; Johnson et al. 2021).

Overall, the anoxic, ferruginous redox state of the early Archean ocean may have impacted total biological productivity by inhibiting biomass remineralization as a source of recycled nutrients (Kipp and Stüeken 2017) and possibly limiting the supply of P and N to primary producers. However, relatively speaking, the biological production of $CH_4$ was probably enhanced by heightened supplies of Ni and Co from mafic rocks. It is conceivable that these conditions are a common feature on other worlds on which biological $O_2$ production has not (yet) evolved or $O_2$ has not been able to accumulate. Mafic (Fe-rich) crustal rocks are prevalent at the surface of planets, and therefore, potential oceans on those planets may also be enriched in dissolved iron. The ferruginous early Archean ocean marks a crucial data point in assessing habitability and biosignature production. Quantitative constraints derived from the trends described above are presented below.

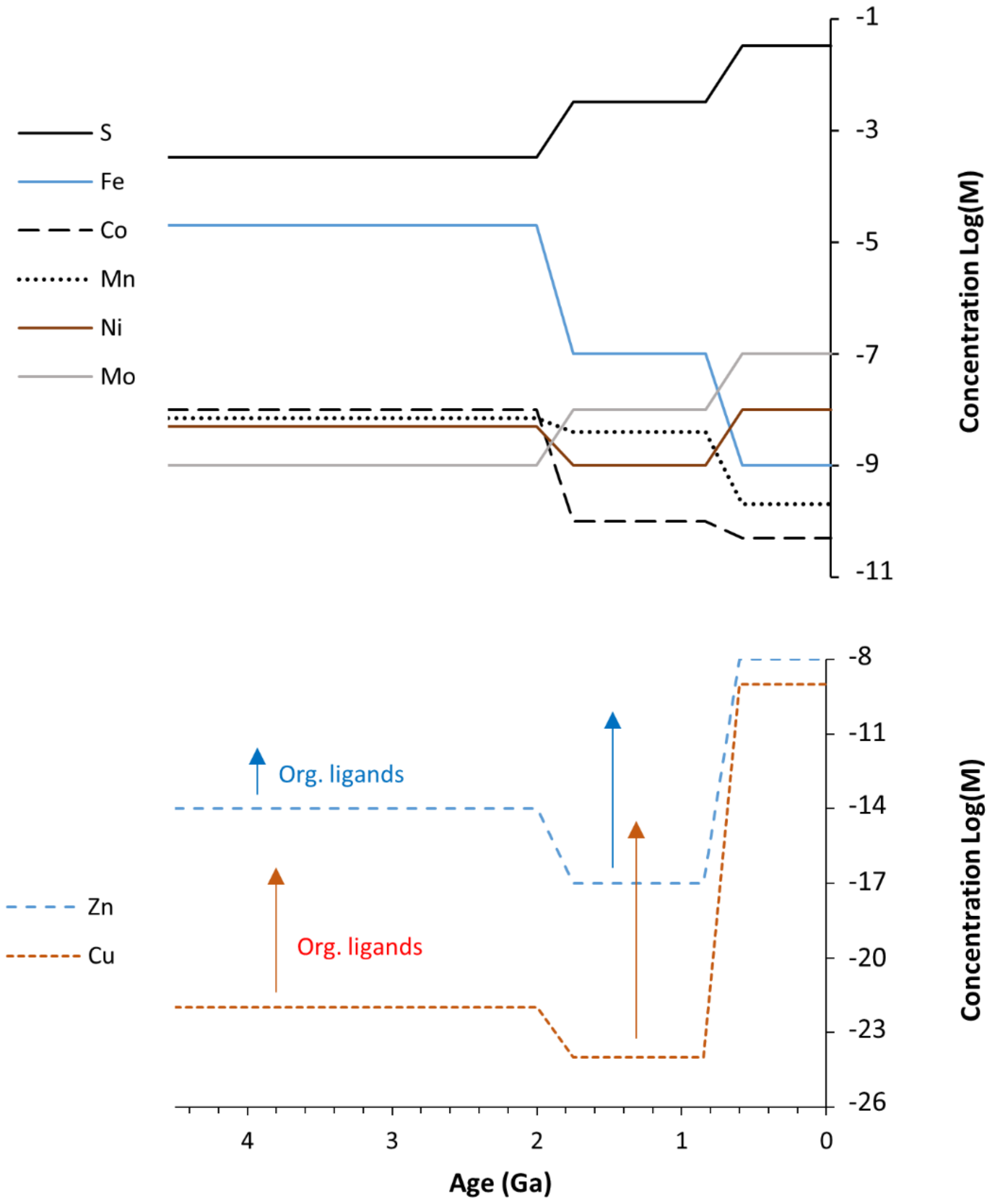

***Figure 6:*** *Modelled concentrations of trace elements through time that are important protein cofactors. The figure is primarily adapted from Saito et al. (2003) with additional incorporated results from other studies (Anbar and Knoll 2002; Zerkle et al. 2005; Anbar 2008; Robbins et al. 2016). Note that Cu (and possibly Zn) solubility could have been elevated above the indicated concentrations in the presence of organic ligands, as indicated by arrows in the lower panel (Stüeken 2020).*

**Microbial life in a redox-stratified world with modern plate tectonics (ca. 3-0.5 Ga)**

***The onset of modern plate tectonics.*** The early Archean Earth was a different planet from today with few emergent land masses, an anoxic atmosphere and ocean, strong volcanic activity, marine nutrient inventories buffered by mafic crust, and possibly a different tectonic style than what we have today. However, from ca. 3 Ga onwards, multiple lines of evidence indicative of plate tectonics with subduction zones, as well as cool and buoyant continental crust, appear in the geologic record (Smithies et al. 2005; Smithies et al. 2007; Condie and Kröner 2008; Shirey and Richardson 2011; Van Kranendonk 2011; Dhuime et al. 2012; Korenaga 2013; Brown and Johnson 2018; Bauer et al. 2020). On the modern Earth, most marine life resides near continental margins, *i.e.*, in the vicinity of riverine nutrient sources and within the photic zone. Hence the establishment of large continental shelves in the Neoarchean would have provided an important habitat space for the biosphere.

The mantle has progressively cooled from the Archean onwards (Herzberg et al. 2010). This cooling contributed to continental emergence via isostasy (Flament et al. 2008) – changes in crustal compositional from more mafic to more felsic were important as well (e.g., Dhuime et al. 2015; Tang et al. 2016) - and may have played a role in the initiation of plate tectonics (O'Neill et al. 2016), though some geodynamical models find that mantle cooling is not a significant factor in promoting plate tectonics (Korenaga 2017; Foley 2018). In any case, the establishment of subduction and mantle cooling promoted volatile recycling into the mantle. Water, in particular, can be more readily regassed into the mantle as mantle temperature declines (Rüpke et al. 2004; Magni et al. 2014; Dong et al. 2021). Xenon isotopes indicate that Earth switched to a state of net water ingassing at the end of the Archean (Parai and Mukhopadhyay 2018).

Volcanism, and therefore volatile outgassing, is affected by both mantle cooling and water ingassing. A cooling mantle lowers volcanism rates due to less extensive melting and less vigorous convection, which causes a lower flux of mantle material into the melting regions at mid-ocean ridges or subduction zones. However, water recycling can counteract this affect, as increasing mantle water content promotes melting by lowering the solidus and decreases mantle viscosity, hence increasing convective vigor (Crowley et al. 2011; Seales et al. 2022). Various models have been constructed of volatile in- and outgassing taking these competing effects into account, finding a range of behaviour from steadily decreasing outgassing rates to outgassing rates that stay relatively constant through this period of Earth's history (McGovern and Schubert 1989; Tajika and Matsui 1992; Fuentes et al. 2019). More efficient water ingassing as the mantle cools implies that the ocean volume has been declining over this period of Earth's history. However, this does not mean that the sea level relative to continents, or freeboard, changed significantly in response. The volume of ocean basins itself is not constant over time due to changes in both continental area and seafloor depth as the mantle evolves (Schubert and Reymer 1985). Korenaga *et al.* (2017) showed that a relatively constant freeboard could be maintained even with a net loss of water from the surface to the mantle. All told, the likely operation of plate tectonics and subduction from the Mesoarchean onwards indicates the presence of significant volcanic outgassing, hydrothermal activity on the seafloor, and deep volatile cycling into the mantle, though the former two would likely have been active even without plate tectonics, as discussed above.

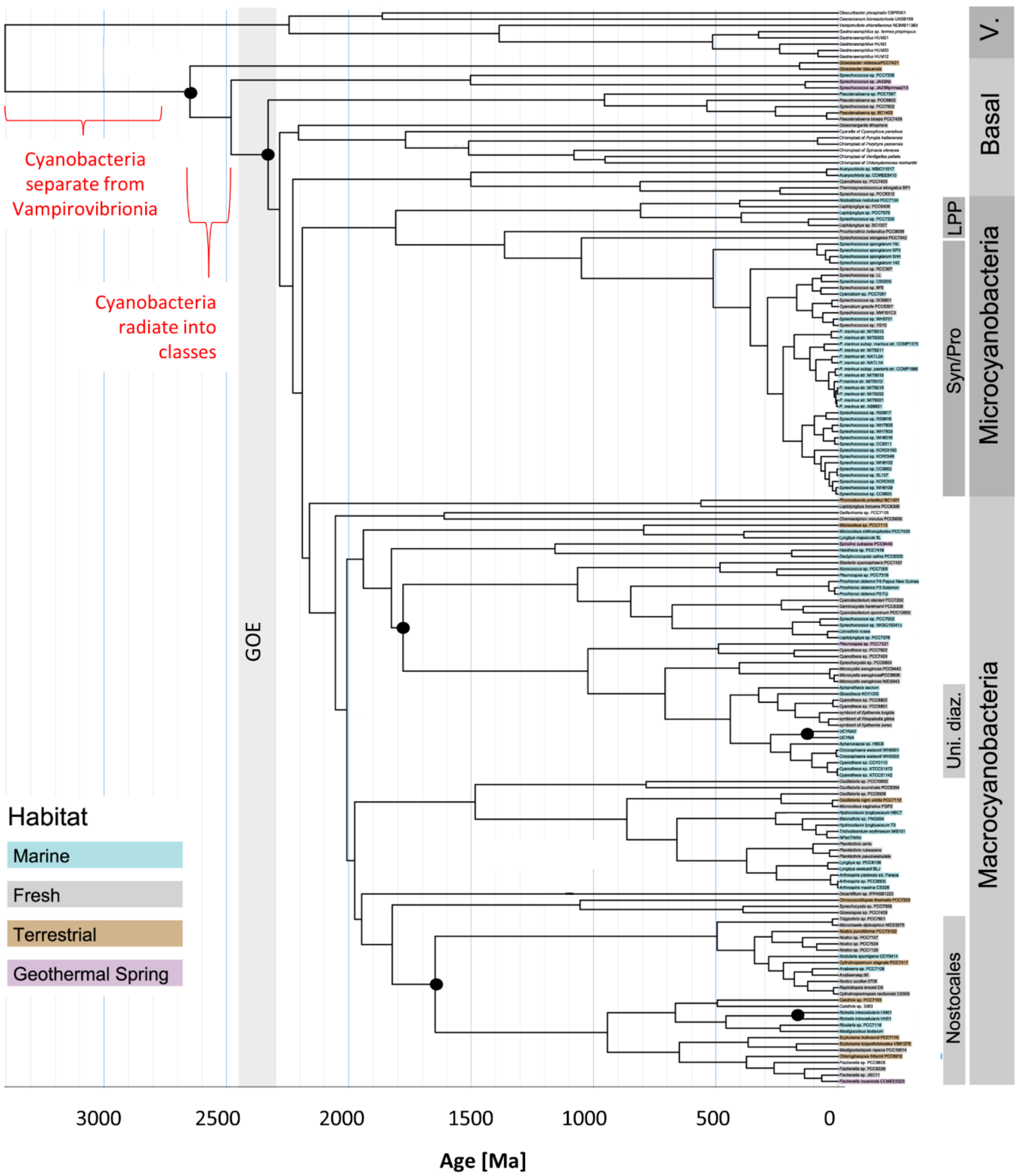

***Figure 7:*** *Phylogeny of cyanobacteria, adapted from Boden et al. (2021) under* *<u>Creative Commons Attribution License CC-BY</u>. According to this reconstruction, the origin of cyanobacteria occurred in the Archean, and diversification took place just prior to the GOE.*

***Oxidation to oxygenation: the onset and consequences of oxygenic photosynthesis.*** Approximately concurrent with the establishment of modern-style plate tectonics, surface environments on Earth began to show the first signs of oxidation (Lyons et al. 2014). By 2.4-2.3 Ga, S-MIF, the key indicator of an anoxic atmosphere, disappeared from the sedimentary rock record (Bekker et al. 2004; Warke et al. 2020; Uveges et al. 2023). Around the same time, paleosols (fossilized soil horizons) started to retain iron in the form of $Fe^{3+}$ minerals (Rye and Holland 1998), detrital pyrite and uraninite (reduced Fe and U minerals) disappeared from river sediments (Johnson et al. 2014), and iron oxide-stained red beds appeared in tidal flat deposits (Eriksson and Cheney 1992). This transition is known as the Great Oxidation Event (GOE, Holland 2002), when atmospheric $O_2$ levels permanently increased to >$10^{-5}$ PAL (Pavlov and Kasting 2002). While some photochemical processes are able to generate $O_2$ gas (Haqq-Misra et al. 2011), by far the major source of free $O_2$ on Earth is oxygenic photosynthesis (reviewed by Catling 2014). Hence the GOE is an expression of biological activity, likely facilitated by a gradual shift in tectonic regime (Van Kranendonk et al. 2012) and possibly progressive biomass burial that removed reducing potential from Earth's surface (Krissansen-Totton et al. 2015). The metabolism was invented in the cyanobacterial phylum of bacteria and marked a turning point in the history of Earth's surface evolution.

The GOE at ca. 2.4-2.3 Ga is the latest possible time when oxygenic photosynthesis could have originated, given the convergence of numerous proxies listed above that indicate the presence of free $O_2$ in terrestrial and shallow-marine environments at that time. While some data and models indeed suggest that the invention of $O_2$-genesis occurred only shortly before the GOE (Johnson et al. 2013; Ward et al. 2016; Shih et al. 2017), other studies indicate that several tens to hundreds of millions of years passed between this biological innovation and the accumulation of $O_2$ in the atmosphere to appreciable levels (see reviews and models by Kurzweil et al. 2013; Lyons et al. 2014; Laakso and Schrag 2017). This time delay is thought to have been required to saturate and/or reduce the flux of various $O_2$ sinks (Zahnle et al. 2013). In other words, reductants in the form of crustal minerals, dissolved ferrous iron in the surface ocean and metamorphic gases first needed to be oxidized before the atmosphere could be oxygenated.

The oldest widely accepted evidence of some form of phototrophy dates back to 3.5 Ga and is preserved in the form of stromatolites and microbially-induced sedimentary structures that are reminiscent of modern photosynthetic microbial mats (Walter et al. 1980; Noffke et al. 2013; Baumgartner et al. 2020b). A larger stromatolite reef occurs in only slightly younger rocks at ca. 3.4 Ga (Allwood et al. 2006). On the modern Earth, stromatolites form when light-sensitive microbial colonies grow upwards towards the Sun. By trapping and binding suspended sediment particles and inducing precipitation of carbonate (typically by $CO_2$ uptake, which raises the pH around autotrophic cells), these microbial structures slowly lithify, often with characteristic laminae (Golubic et al. 2000). Today, most stromatolites are dominated by the primary producers cyanobacteria (e.g., Neilan et al. 2002). However, this was not necessarily the case at 3.5 Ga. It is conceivable that photosynthetic organisms at that time were anoxygenic, *i.e.*, oxidizing sulfide, $H_2$, or ferrous iron instead of $H_2O$ (Baumgartner et al. 2020b), and evidence of free $O_2$ has not been documented from these rocks. Some phylogenetic reconstructions suggest that anoxygenic photosynthesis pre-dates oxygenic photosynthesis (Raymond and Blankenship 2008), though this aspect is debated (Cardona et al. 2015). In any case, the presence of stromatolites in the Early Archean is perhaps evidence that life learned relatively early how to harvest photochemical energy from the Sun (see also Cardona et al. 2017; Cardona et al. 2019). Early life on Earth may therefore not solely have been dependent on a supply of electrochemical energy from abiotic sources such as volcanism, UV-photolysis or lightning. Instead, a flux of reductants and stellar photons were perhaps sufficient for phototrophic cells to generate electrochemical energy internally.

Liquid water is the most abundant reductant on the planet, and hence the appearance of oxygenic photosynthesis - where $H_2O$ is oxidized to $O_2$ - is thought to have spurred primary productivity by orders of magnitude (Kharecha et al. 2005; Lyons et al. 2014). The first geochemical evidence for $O_2$ in the surface ocean dates back to 3 Ga, where molybdenum isotopes display a covariance with Mn concentrations in sedimentary iron formations that is most parsimoniously explained by the former presence of Mn-oxides in those rocks (Planavsky et al. 2014). Hence $O_2$ production may have started at or shortly before 3 Ga, which is also consistent with some phylogenetic reconstructions (Sánchez-Baracaldo and Cardona 2020; Boden et al. 2021, Fig. 7) (though see alternative interpretatiosn by Shih et al. 2017). However, the near persistence of S-MIF throughout the Archean (with a possible gap around 3 Ga) indicates that $O_2$ levels in the global atmosphere remained below $10^{-5}$ PAL, meaning that $O_2$ would have been a trace gas and likely undetectable by remote techniques for perhaps the entire Archean (though the Mesoarchean interval with muted MIF around 3 Ga remains to be explained, possibly by the presence of $O_2$ or by an atmospheric haze at that time (Domagal-Goldman et al. 2008)). As noted earlier, the main reason for lack of $O_2$ accumulation despite ongoing production was the high abundance of reductants that first needed to be saturated (oxidized). This is an important lesson for observations of habitable exoplanets, because the atmospheric signal of $O_2$-producers may be undetectable if it is swamped by the flux of biotic and abiotic reductants from the surface of the planet.

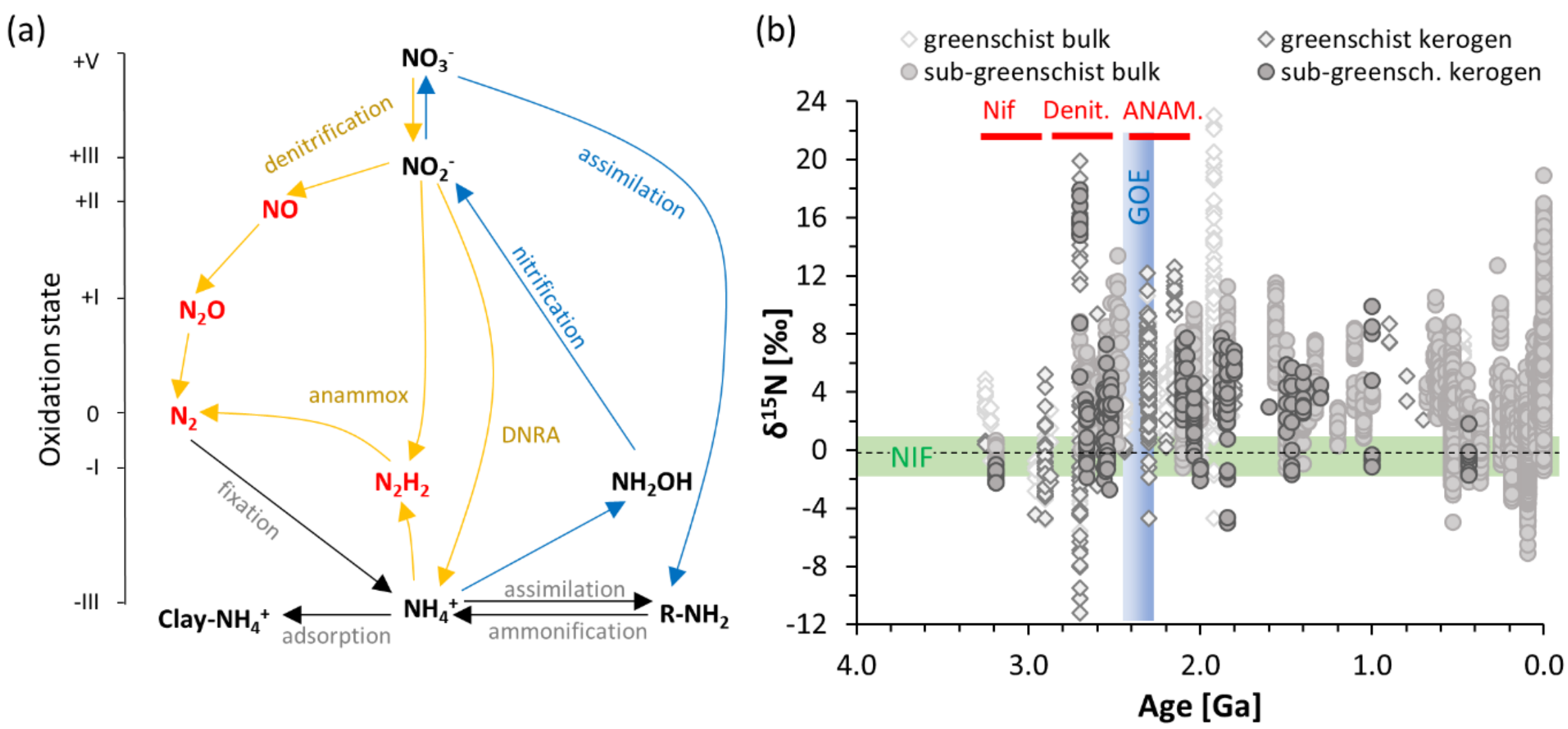

***Figure 8:*** *(a) Schematic of the biological nitrogen cycle. Black = entirely anoxic pathways; orange = pathways occurring under anoxic or suboxic conditions; blue = pathways that typically indicate the presence of oxic conditions. Species highlighted in red may occur as gases in the atmosphere. (b) The sedimentary $\delta^{15}N$ record through time for rocks up to greenschist metamorphic facies. Large nitrogen isotope fractionations are associated with redox reactions involving nitrification and denitrification, and therefore the increasing spread in $\delta^{15}N$ seen from ~2.8 Ga onwards may reflect the onset of these metabolisms. Red bars = phylogenetic constraints on the first appearance of $N_2$ fixation (Nif), denitrification (Denit.) and anammox (ANAM.) (Parsons et al. 2020; Liao et al. 2022). Nitrogen isotope database taken from Stüeken et al. (2016) with recent updates from the literature (Zerkle et al. 2017; Kipp et al. 2018; Koehler et al. 2018; Luo et al. 2018; Koehler et al. 2019; Mettam et al. 2019; Ossa Ossa et al. 2019; Yang et al. 2019; Gilleaudeau et al. 2020; Hodgskiss et al. 2020; Wang et al. 2020; Stüeken et al. 2021b; Stüeken and Prave 2022; Stüeken et al. 2022). GOE = Great Oxidation Event.*

Indirect geochemical evidence of oxic surface waters in the Archean ocean becomes more widespread between 2.7-2.5 Ga (Buick 1992; Wille et al. 2007; Kendall et al. 2010; Flannery and Walter 2012; Riding et al. 2014; Koehler et al. 2018; Wilmeth et al. 2022). Some of this evidence includes nitrogen isotope data (Fig. 8) indicative of a redox-active nitrogen cycle where nitrate was able to accumulate at least transiently (Garvin et al. 2009; Godfrey and Falkowski 2009; Koehler et al. 2018). The reduction of nitrate at the interface between oxic surface waters and anoxic deep waters (where it commonly occurs today in anoxic setting, see review by Pajares and Ramos 2019) may have led to the production of $N_2O$ gas (see below). Phylogenetic data suggest that biological nitrate reduction to nitrite emerged in the Neoarchean (Parsons et al. 2020), but nitrite reduction only emerged in the late Proterozoic. It is likely that high levels of ferrous iron abiotically reduced nitrite, where NO and $N_2O$ would have been major products (Stanton et al. 2018; Buessecker et al. 2022). Hence increasing fluxes of $N_2O$ and possibly NO from the ocean into the atmosphere could have created a novel biosignature from the Neoarchean onwards.

Although the GOE marks the time when many surface environments became oxidized, atmospheric $O_2$ levels may have remained as low as 0.1-1% throughout the Proterozoic (Lyons et al. 2014), possibly due to a continuously high flux of reduced iron from hydrothermal sources in the ocean that declined only slowly with mantle cooling (Planavsky et al. 2018a; Wang et al. 2022a), though slightly higher levels up to 10% are achieved in some models (Laakso and Schrag 2017; Crockford et al. 2018; Fakhraee et al. 2019). In any case, the deep ocean appears to have remained largely anoxic and buffered by ferrous iron (Reinhard et al. 2013; Sperling et al. 2015). Oxygenation of the deep ocean may have occurred only periodically (Yang et al. 2017; Planavsky et al. 2018b; Stüeken et al. 2021b). Under these conditions, it is also possible that productivity and hence $O_2$ production were nutrient-limited (Crockford et al. 2018; Laakso and Schrag 2019).

Modern-like $O_2$ levels were probably only reached in the Neoproterozoic or early Phanerozoic (0.8-0.5 Ga) (Och and Shields-Zhou 2012), and possibly as late as the Devonian rise of land plants at 0.35 Ga (Lenton et al. 2016; Krause et al. 2018). This delay probably also held back the expansion of complex life, which is largely dependent on $O_2$ as an electrochemical energy source (Reinhard et al. 2016). Understanding biogeochemical cycles and feedback loops on this redox-stratified Proterozoic Earth may therefore be important for assessing the likelihood of complex organisms arising on other worlds.

***Ocean redox stratification.*** Crustal differentiation and changing tectonic regimes in the Neoarchean led to a decline in the availability of (ultra-)mafic crust, which is thought to have resulted in reduced Ni availability in seawater (Konhauser et al. 2009; Konhauser et al. 2015). Nickel is a metal cofactor in methanogenesis (Diekert et al. 1981), perhaps one of the oldest metabolisms on Earth (Woese 1987), and so the decline in Ni availability may have caused a methanogen famine (Konhauser et al. 2009; Konhauser et al. 2015). In other words, Ni-based methanogenesis may have been elevated in the earlier Archean when Ni supplies were higher. Similar to Ni, Co is an important cofactor in deeply rooted metabolic pathways (hydrogenotrophic acetogenesis and methanogenesis, see Gottschalk and Thauer 2001; Ragsdale and Pierce 2008), whose availability decreased with the oxygenation of the ocean (Moore et al. 2018). Manganese is an essential element in the oxygen evolving complex (OEC) of all oxygenic photosynthesizers (Ghanotakis and Yocum 1990; Roelofs et al. 1996). The ability of the OEC and photosystem II to split water and take the protons and electrons from the $H_2O$ molecule to reduce the carbon in $CO_2$ is a crucial biological innovation, because it allows oxygenic photosynthesizers to use highly abundant water as an electron source. Manganese was more available in the Archean when geochemical evidence indicates that oxygenic photosynthesis evolved (Crowe et al. 2013; Planavsky et al. 2014; Cardona et al. 2019) (see also above). The rise of oxygen has led to decreasing Mn availability through time (Robbins et al. 2023).

Increasing $O_2$ levels towards the end of the Archean and across the GOE likely impacted nutrient availability in two main ways (see also Saito et al. 2003; Hao et al. 2017) (Fig. 9). First, elements or species that are more soluble in the presence of $O_2$ would have become more bioavailable in oxic surface waters at this time. Key examples probably included nitrate (see above) as well as Mo in the form of molybdate (Scott et al. 2008; Reinhard et al. 2013). Evidence for oxic surface waters during the Proterozoic is preserved in the form of elevated iodate concentrations in shallow marine carbonates (Hardisty et al. 2017). Second, increasing $O_2$ in the atmosphere would have stimulated the sulfur cycle through the onset of oxidative weathering on land with indirect effects on the solubility of sulfur-sensitive metals. In the Archean, isotopic data suggest that sulfate levels were <200 μM (less than one hundredth of present levels, Habicht et al. 2002) and possibly below 2.5 μM (Crowe et al. 2014), which resulted in supressed sulfate reduction and a carbon cycle dominated by methanogenesis. The major source of sulfur to the ocean was volcanic outgassing. Following the GOE, sulfate levels probably increased to 10 mM as indicated by the first occurrence of sulfate evaporites (Blättler et al. 2018). Lower levels down to 100 μM have been proposed for the later Proterozoic (Fakhraee et al. 2019). In any case, the elevated flux of sulfate to the deep ocean, where conditions continued to be anoxic, resulted in the production of hydrogen sulfide ($H_2S$) through biological sulfate reduction. Sulfide rapidly reacts with ferrous iron, forming pyrite in sediments. Where sulfide dominates over iron in the water column, conditions are referred to as euxinic. While it was once thought that such euxinic conditions extended over the entire Proterozoic deep ocean (Canfield 1998), more recent studies on iron speciation have shown that euxina was limited to productive continental margins, akin to modern oxygen minimum zones (Planavsky et al. 2011; Poulton and Canfield 2011; Lyons et al. 2014). Quantitative models suggest that <10 % of the deep ocean was overlain by euxinic waters (Reinhard et al. 2013). Nevertheless, the residence time of transition metals in the ocean was significantly impacted (Fig. 6). In particular Cu and Zn would have been scavenged more rapidly from the water column (Saito et al. 2003), unless stabilized by organic ligands (Stüeken 2020), and also the availability of Mo was overall reduced (Reinhard et al. 2013).

Copper is a crucial metal cofactor for aerobic respiration (Klinman 1996) and denitrification (Delwiche and Bryan 1976), but unlike the marine concentrations of many metal cofactors described above, the availability of Cu probably increased with ocean oxygenation in the Neoproterozoic or Phanerozoic thus supporting aerobic metabolisms (Saito et al. 2003; Ciscato et al. 2019). Similarly, the availability of Zn increased during roughly the same time period (Scott et al. 2012), which allowed for greater utilization by eukaryotes, including DNA binding zinc-finger proteins (Dupont et al. 2006; Dupont et al. 2010). In other words, both Cu and Zn were scarce and possibly limiting for certain metabolisms on the fully anoxic early Archean Earth.

Persistent anoxia in the deep ocean was probably promoted by a combination of continued influx of $Fe^{2+}$ from hydrothermal activity and a generally warm climate, which reduced oxygen solubility (Knauth 2005). Some studies propose that anoxic conditions led to sedimentary phosphate release, and the enhanced nutrient supply fuelled high biological productivity and oxygen demand resulting in further oxygen depletion and sulfide build-up via sulfate reduction (Meyer and Kump 2008). This may be consistent with the appearance of phosphorites in parts of the Proterozoic (Papineau 2010). However, as noted above, the behavior of P in the Precambrian ocean is contested (Derry 2015; Reinhard et al. 2017b; Laakso and Schrag 2019; Brady et al. 2022; Ingalls et al. 2022; Kipp 2022), and it is also possible that anoxia was maintained because low nutrient availability limited biological $O_2$ production (Crockford et al. 2018; Laakso and Schrag 2019; Ozaki et al. 2019; Hodgskiss et al. 2020). Remaining uncertainties about the causes of marine redox stratification during the Proterozoic make it difficult to extrapolate this scenario to other inhabited planets. However, given the longevity of this state in Earth's history, it appears plausible that similar conditions could arise elsewhere, provided that $O_2$ production emerges. Hence understanding the production of biosignatures on a Proterozoic-like planet is crucial in our search for life elsewhere.

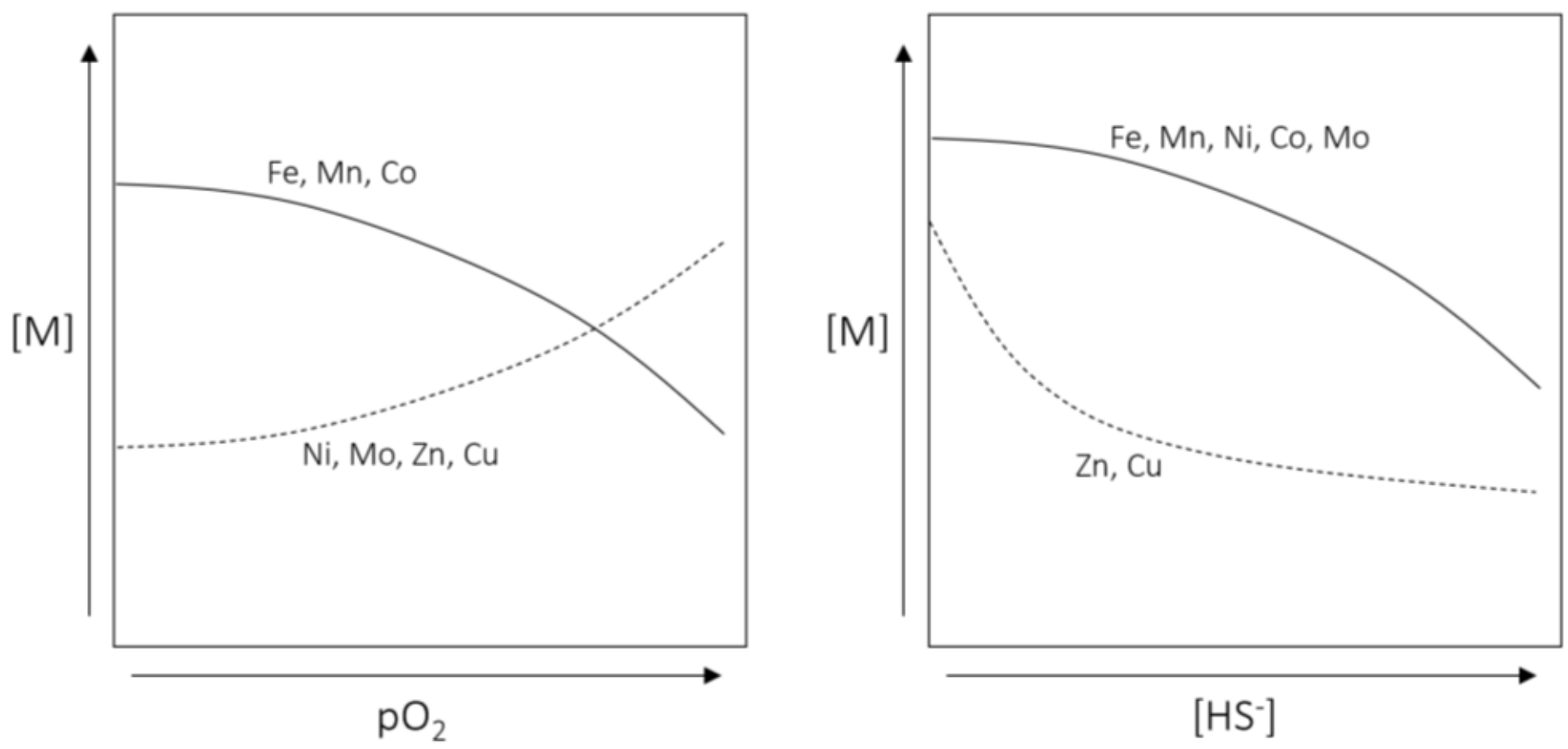

***Figure 9:*** *Trace metal concentration responses to increasing $pO_2$ and [$HS^-$]. The stronger impact of sulfide complexation on the solubility of Zn and Cu compared to other metals is based on differences in solubility constants (Saito et al. 2003).*

## QUANTITATIVE CONSTRAINTS ON ATMOSPHERIC BIOSIGNATURES THROUGH TIME

Armed with this framework of Earth's geobiological evolution over the past four billion years, we can now place constraints on the types of biosignatures that microbial biospheres are able to generate and that may be detectable on exoplanets with similar environmental conditions. We have detected 5000+ exoplanets, which is just a tiny fraction of the planets in our galaxy (Christiansen 2022). The nearest world to us, Proxima Centauri b, is 4.2 light years away (Anglada-Escudé et al. 2016). It is just one of a couple dozen of known planets that are roughly Earth-sized and in the Habitable Zone of their stars where surface liquid water and life may be possible (Hill et al. 2023). Whether any of these worlds host life is among the most fundamental questions of science. However, all of these worlds are sufficiently distant that we will not be able to search for signs life in the same ways as we have in Earth's earliest geological record, on present-day Mars, and the icy moons of our Solar System.

We must instead remotely recognize the presence of alien biospheres and characterize their biogeochemical cycles in planetary spectra obtained with large ground- and space-based telescopes (e.g., Fujii et al. 2018; Schwieterman et al. 2018). These telescopes can probe atmospheric composition by detecting absorption features associated with specific gases. We might also be able to recognize global-scale surface features, including light interaction with photosynthetic pigments (Seager et al. 2005) and 'glint' arising from specular reflection of light by a liquid ocean (Robinson et al. 2010).

Unfortunately, key details regarding marine environments will be inaccessible with telescopes (Olson et al. 2020). For example, neither the oxygen landscape of the ocean nor availability of bioessential nutrients can be directly observed. Moreover, biogenic gases produced by marine microbes may be consumed by other microbial metabolisms before entering the atmosphere, and biogenic gases do not necessarily accumulate in the atmosphere or imprint their presence in planetary spectra (Reinhard et al. 2017a). The detectability of biosignatures is also strongly influenced by instrument design and viewing geometries. The wealth of information contained within Earth's geochemical and fossil records thus does not always directly inform the signatures of Earth life that would be detectable to alien observers of our planet. It is nonetheless an important starting point. Unravelling the details of Earth's complex biogeochemical history and its relationship with remotely observable spectral signals is an important consideration for instrument design and our own search for life in the Universe (Reinhard et al., 2019).

**Oxygen ($O_2$)**

Oxygen is by far the most commonly discussed remote biosignature (Meadows et al. 2018; Schwieterman et al. 2018). Earth's oxygen-rich atmosphere that we depend on is itself the consequence of life, specifically oxygenic photosynthesis, which probably dates back to >3 Ga (see above). Hence detecting $O_2$ in an exoplanet atmosphere could suggest the presence of life beyond Earth. However, although $O_2$ is a glaring signal of life on our planet, the presence of $O_2$ does not uniquely indicate life. This is because there are several non-biological ways to produce $O_2$, potentially leading to abiotic signals that mimic life (biosignature "false positives"). Ways of abiotic oxygen production include $H_2O$ photolysis and subsequent H escape to space, driving atmospheric oxidation. This scenario is particularly likely to play out on M dwarf planets that experience a hostile UV environment during the pre-main sequence evolution of their star, possibly leading to planetary desiccation and very high levels of abiotic oxygen (e.g., Luger and Barnes 2015). Photolysis of $CO_2$ can also lead to abiotic $O_2$ accumulation on M-dwarf planets with $H_2O$-depleted, $CO_2$-dominated atmospheres (Gao et al. 2015). A scenario for abiotic $O_2$ on planets orbiting Sun-like stars is the lack an effective "cold trap." $H_2O$ condenses as rising air expands and cools, effecting trapping $H_2O$ near Earth's surface. However, planets with low inventories of non-condensing gases such as $N_2$ allow $H_2O$ to enter the upper atmosphere, enhancing $H_2O$ photolysis, subsequent H escape, and planetary oxidation (Wordsworth and Pierrehumbert 2014). High obliquity (tilt of the rotational axis with respect to the orbital plane) may also lead to weakening of the cold trap (Kang 2019). Fortunately, these scenarios can be recognized as non-biological with appropriate context. For example, $CO_2$ photolysis produces CO in addition to $O_2$; high surface pressure can help rule out scenarios involving a weak cold trap; and many bars of $O_2$ or a lack of $H_2O$ may be incompatible with life (Meadows et al. 2018). Oxygen must therefore be considered in its broader stellar, orbital, and planetary context, considering both the presence and absence of other chemical species.

Earth's history highlights the possibility of "false negatives" for life as well (Reinhard et al. 2017a), a challenge that may be more difficult to overcome. As discussed above, oxygen was much lower than today for the vast majority of Earth's history, even long after the onset of biological oxygen production during photosynthesis due to continued influx of reductants from the crust and hydrothermal sources. Oxygen was limited to "oxygen oases" in the Archean surface ocean for up to 500 million years before the GOE and atmospheric $pO_2$ did not approach near modern levels for another 2 billion years after that (Olson et al. 2018). Prior to the GOE, oxygenic photosynthesis would have been undetectable to a remote observer based on known Earth-based technology. For the next 2 billion years of the Proterozoic, $pO_2$ (0.1-10% modern levels, e.g. Laakso and Schrag 2017; Wang et al. 2022b) was elevated relative to the Archean but still lower than today, as the ocean continued to experience a high influx of Fe(II) from hydrothermal sources and a lack of nutrients potentially limited oxygenic photosynthesis (see above). Detection of $O_2$ at the high end of estimated levels may be feasible with a future direct imaging mission but will nonetheless depend strongly on instrument design and favorable viewing geometries (Young et al. 2023a). Oxygen levels at the lower end of estimates will likely be inaccessible remotely (Robinson and Reinhard 2018). We thus must be mindful that Earth-like biospheres on exoplanets may lack detectable $O_2$.

Fortunately, even low levels of oxygen may lead to photochemical production of $O_3$. Ozone absorbs strongly in the UV and mid-IR, and it is much more readily detectable than $O_2$. Ozone may therefore be our most reliable indicator of Proterozoic-like biospheres (Reinhard et al. 2017a; Reinhard et al. 2019), even though it is not a direct biological product. Ultimately, understanding the relationships between life, marine habitats, and atmospheric chemistry is essential to interpreting detections (or non-detections) of oxygen in exoplanet atmospheres. The absence of detectable atmospheric oxygen does not preclude aerobic ecosystems (Olson et al. 2013), but it may preclude animal-grade complexity (Catling et al. 2005). On the other hand, detection of $O_2$ or its photochemical product $O_3$ does not necessarily imply a persistently well-oxygenated ocean capable of supporting animal-grade complexity (Reinhard and Planavsky 2022). Nevertheless, $O_2$ remains an important target for next-generation telescopes in the search for life despite these challenges.

**Methane ($CH_4$)**

Earth's $CH_4$ is overwhelmingly biological, both today and in the past. Moreover, methanogenesis is an ancient metabolism on Earth, dating back to 3.5 Ga (see above). As noted above, it is formed from $CO_2$ and $H_2$, which are generated by volcanic and hydrothermal processes that are likely common in the Universe. Methane is thus widely considered a potential biosignature. However, several abiotic sources of $CH_4$ are also known. Volcanic outgassing, serpentinization and impacts can all produce $CH_4$, albeit at low rates compared to life on Earth (Thompson et al. 2022). $CH_4$, not unlike $O_2$, thus requires thorough consideration of its context. For example, the co-existence of $CO_2$ (C in its most oxidized form) and $CH_4$ (C in its most reduced form) represents chemical disequilibrium, which is most readily explained by life in the absence of their redox intermediate, CO (Krissansen-Totton et al. 2016;

Thompson et al. 2022). Abiotic $CH_4$ is common in the solar system, and equilibrium chemistry on some exoplanets may produce co-existing $CH_4$ and $CO_2$—but all these scenarios involve production of CO alongside $CH_4$. Co-existence of $CH_4$ and $CO_2$, without CO, thus implies sustained surface fluxes of both endmember redox species and is a robust signature of life (Krissansen-Totton et al. 2018a).

It is widely believed that lower oxygen in the ancient ocean and atmosphere led to higher levels of atmospheric methane (e.g., Kasting 2005). The reasons are three-fold. First, lower levels of $O_2$ would have directly limited aerobic respiration and indirectly limited anaerobic respiration through the scarcity of alternate electron acceptors such as $NO_3^-$ or $SO_4^{2-}$ that are most readily produced in the presence of $O_2$. This limitation increases the ecological importance of less energetic metabolisms, such as methanogenesis. Second, the scarcity of electron acceptors further limits biological consumption of $CH_4$, increasing the fraction of biogenic $CH_4$ that actually reaches the atmosphere. Finally, an atmosphere that was less oxidizing could be more favorable to $CH_4$ accumulation. However, quantifying exactly how atmospheric $CH_4$ has varied through time nonetheless remains challenging.

Whereas oxygen is reactive and leaves its fingerprints all over the geochemical record (reviewed above), $CH_4$ leaves a more subtle signal. As noted earlier, the earliest geochemical hints of biogeochemical methane cycling included isotopically depleted organics and $CH_4$ gas inclusions in Archean sediments at 3.5-3.4 Ga (Ueno et al. 2006; Flannery et al. 2018). This signature becomes more conspicuous in the Neoarchean around 2.7 Ga (e.g., Eigenbrode and Freeman 2006). Conventional interpretation of these Neoarchean data implies biological assimilation of methane, which is typically isotopically depleted relative to other C substrates. This signature reflects $CH_4$ consumption (*i.e.,* methanotrophy) from a methanogenic source, and although it could be compatible with high concentrations of $CH_4$ it only requires $CH_4$ production, with or without accumulation. An alternative interpretation of isotopically light organics in the Archean is that these organics were produced photochemically and subsequently deposited from the atmosphere, implying a sufficiently $CH_4$-rich atmosphere to trigger hydrocarbon polymerization in an organic haze (Pavlov et al. 2001). This interpretation may also find support in nuanced S isotope patterns that may arise from fluctuations in UV attenuation by an organic haze (Zerkle et al. 2012; Izon et al. 2017). Haze forms for $CH_4$:$CO_2$ ratios > 0.1 (e.g., Zerkle et al. 2012). Reasonable estimates of Archean $pCO_2$ imply $CH_4$ levels on the order of ~1000 ppm (Izon et al. 2017), which would be detectable with James Webb Space Telescope (JWST) and next-generation telescopes (Krissansen-Totton et al. 2018b). The development of such a haze may affect our ability to characterize some planetary features, but such a haze may itself constitute a biosignature (Arney et al. 2018). It is worth noting that abiotic hazes are common in the outer solar system  and may be common among close-in exoplanets with radii larger than Earth but smaller than Neptune ("sub-neptunes", Bergin et al. 2023; Gao et al. 2023). The likelihood of abiotic hazes highlights the importance of interpreting hazes in their planetary context, not unlike other biosignatures.

Atmospheric $CH_4$ is thought to have plummeted in association with oxygenation during the GOE, but it remains debated whether such a decline was a cause (decreased O2 sink, Konhauser et al. 2009) or consequence (less methanogeneis, more methanotrophy, Pavlov et al. 2003). In either scenario, loss of greenhouse warming by $CH_4$ appears to have contributed to global glaciation in the Paleoproterozoic (Kirschvink et al. 2000; Kopp et al. 2005). The establishment of an $O_3$ layer following the GOE may have increased the chemical lifetime of $CH_4$ (Goldblatt et al. 2006), leading to at least partial recovery of $pCH_4$ if similar biogenic fluxes were sustained (Claire et al. 2006). In weakly oxygenated atmospheres, $pCH_4$ actually increases with $pO_2$ due to stronger UV shielding by $O_3$ (Fig. 10).

Higher $CH_4$ fluxes than today would be anticipated from the redox-stratified Proterozoic ocean (see above). However, increased $SO_4^{2-}$ availability arising from oxidative weathering of the continents may have severely curtailed $CH_4$ production and preservation. Anaerobic respiration coupling organic carbon oxidation to $SO_4^{2-}$ reduction is a more energetic metabolism than methanogenesis, limiting $CH_4$ production in the presence of $SO_4^{2-}$. At the same time, anaerobic methanotrophy that couples $SO_4^{2-}$ reduction to $CH_4$ oxidation is an efficient sink for biogenic $CH_4$. As a result, Proterozoic $pCH_4$ may have only been ~10x higher than at present-day (Olson et al. 2016; Reinhard et al. 2020) (Fig. 11), much lower than early estimates of 100x present day levels (Pavlov et al. 2003; Kasting 2005).

Ironically, the decline of $CH_4$ biosignatures in this case is the consequence of life. One microbe's waste is another's food, and a perfectly efficient biosphere would not produce any net $O_2$ or $CH_4$ for remote observers to detect. Accumulation of these gases in the atmospheres requires inefficiency such as organic carbon burial or slow biological destruction of metabolic waste products in the ocean relative to the rate at which they escape to the atmosphere. The strong sensitivity of $pCH_4$ to marine $SO_4^{2-}$ is a challenge for remotely characterizing biospheres because the $SO_4^{2-}$ ion does not exchange with the atmosphere like gases such as $O_2$ and $CH_4$, and $SO_4^{2-}$ does not come to equilibrium with an atmospheric phase. It is thus not possible to infer ocean $SO_4^{2-}$ levels and their consequences for marine $CH_4$ production based on atmospheric chemistry.

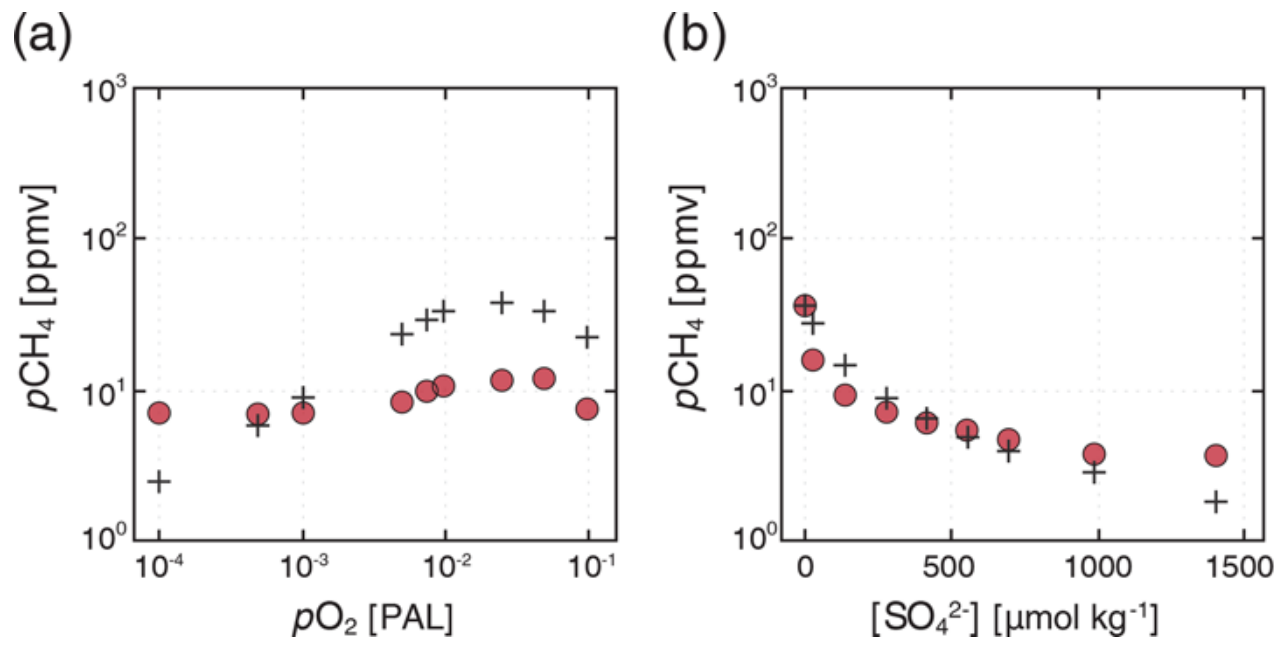

***Figure 10:*** *Sensitivity of atmospheric methane to atmospheric oxygen (left) and oceanic sulfate (right). Black crosses denote simulations from Olson et al., 2016 and red circles are from Reinhard et al. (2020). $pCH_4$ responds non-linearly to $pO_2$ due to UV attenuation by $O_3$ (Claire et al. 2006; Goldblatt et al. 2006), and $pCH_4$ drops off rapidly with increasing $SO_4^{2-}$ (Olson et al. 2016). This figure modified from Reinhard et al. (2020) under Creative Commons Attribution License CC-BY.*

The downward revision of Proterozoic $pCH_4$ somewhat reduces the appeal of $CH_4$ as a biosignature for Proterozoic-like biospheres around Sun-like stars (Reinhard et al. 2017a; Olson et al. 2018), but higher $CH_4$ levels would be expected on planets with similarly productive biospheres orbiting cooler M dwarfs because these worlds receiver fewer UV photons (Segura et al. 2005). Moreover, as a greenhouse gas, $CH_4$ has a number of spectral features in the IR and is among the most accessible biosignatures in the short-term with JWST (Krissansen-Totton et al. 2018b), which is unlikely to detect even modern levels of $O_2$ (Fauchez et al. 2020).

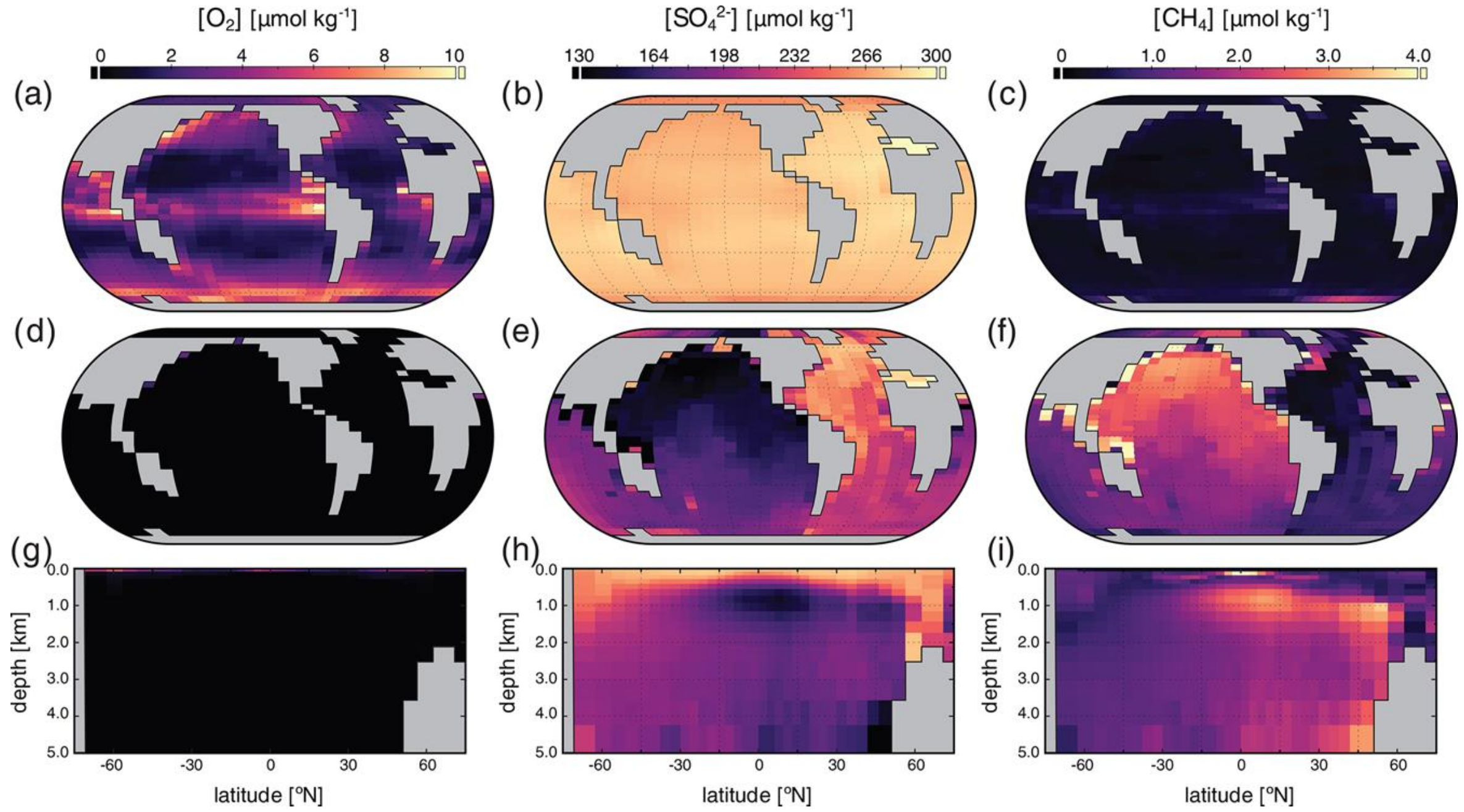

***Figure 11:*** *Simulated oxygen (left; a,d,g), sulfate (middle; b,e,h), and methane (right; c,f,i) concentrations in a Proterozoic-like ocean with $10^{-3}$ PAL $O_2$. Maps show each species in surface seawater (top; a-c) and benthic seawater (middle; d-f). Also shown are zonal average vertical profiles (bottom; g-i). Oxygen is limited to oxygen oases in surface seawater, leading to extensive $SO_4^{2-}$ reduction and local $SO_4^{2-}$ scarcity at depth, especially underlying productive regions of the surface ocean and in the oldest bottom waters. Where $SO_4^{2-}$ concentrations are low, methanogenesis leads to $CH_4$ accumulation, but $CH_4$ must evade oxidation by both $SO_4^{2-}$ in the ocean interior and $O_2$ in surface waters before it may enter the atmosphere. This figure is reproduced from Reinhard et al. (2020) under Creative Commons Attribution License CC-BY.*

## Nitrous oxide ($N_2O$)

Nitrous oxide ($N_2O$) is another popular remote biosignature candidate (e.g., Rauer et al. 2011; Schwieterman et al. 2018). $N_2O$ is produced as a waste product of incomplete denitrification in redox-stratified environments. Denitrification involves the reduction of $NO_3^-$ to $N_2$ via several intermediate species, including $N_2O$. Some $N_2O$ may escape full reduction to $N_2$ ('partial' or 'incomplete' denitrification) and enter the atmosphere. As a greenhouse gas, $N_2O$ has several spectral features in the IR that could fingerprint life at sufficiently high levels with JWST (Schwieterman et al. 2022). Denitrification is limited to $O_2$-depleted environments today. Likewise, denitrification would have been limited on a broadly anoxic Archean Earth because nitrate would have been scarce outside of oxygen oases (Godfrey and Falkowski 2009). However, the redox-stratified ocean of the Proterozoic would provide optimal conditions for $N_2O$ production (see above) (Fig. 12). There would have been enough $O_2$ in Proterozoic surface waters to stabilize $NO_3^-$, but the widely anoxic deep ocean would have allowed extensive $NO_3^-$ reduction (Fennel et al. 2005). As noted above, geochemical evidence for the presence of nitrate in surface waters goes back to approximately 2.7 Ga and becomes more established after the GOE (Godfrey and Falkowski 2009) (Fig. 8).

Unlike $CH_4$, $N_2O$ production would not be strongly curtailed by $SO_4^{2-}$ accumulation in the ocean because denitrification is more energetically favorable than $SO_4^{2-}$ reduction. Sulfide arising from $SO_4^{2-}$ reduction may even enhance global $N_2O$ production if local sulfidic conditions limited the availability of metal co-factors involved in the reduction of $N_2O$ to $N_2$ (Buick 2007b; Roberson et al. 2011).

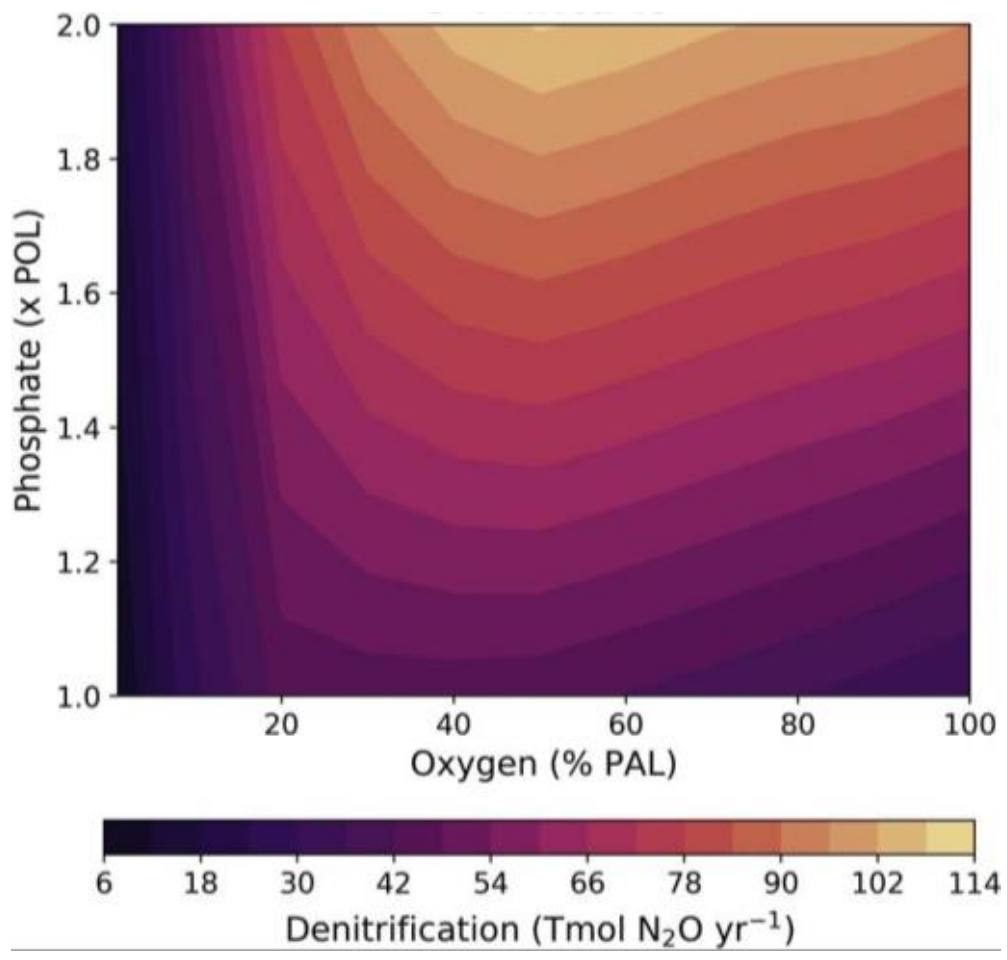

***Figure 12**: Denitrification rates (i.e., maximum $N_2O$ production fluxes; contours) as a function of atmospheric oxygen (x-axis) and oceanic phosphate (y-axis). Dentrification rates are optimized for intermediate $pO_2$. Denitrification is further sensitive to nutrient availability, which dictates the flux of particulate organics available to denitrifiers. Figure reproduced from Schwieterman et al. (2022) under Creative Commons Attribution License CC-BY.*

A recent study showed Proterozoic-like planets orbiting Sun-like stars may accumulate 5-50 ppm $N_2O$ (compared to 330 ppb today) if a significant fraction of the total denitrification flux from the ocean is released as $N_2O$ (Schwieterman et al. 2022). Around M or K dwarf stars, similar early Earth-like fluxes may lead to even higher $pN_2O$—perhaps a 1000+ ppm—owing to muted photochemical destruction, further increasing its appeal as a potential biosignature for weakly oxygenated worlds (Segura et al. 2005; Schwieterman et al. 2022). $N_2O$ may be less vulnerable to non-biological false positives than $CH_4$ and $O_2$. As mentioned above, reduction of nitrite by Fe(II) can produce $N_2O$ without direct involvement of life (Stanton et al. 2018; Buessecker et al. 2022). However, the oxidized N species required for this process are an indirect consequence of oxygenic photosynthesis. A scenario where an abiotically oxygenated atmosphere leads to oxidized N that is then reduced by Fe is unlikely because known pathways to abiotic $O_2$ are likely incompatible with an Fe(II)-rich ocean (Schwieterman et al. 2022). $N_2O$ arising from this abiotic pathway is therefore a potential biosignature for exoplanet atmospheres, much like $O_3$ is a potential biosignature despite its abiotic production in the atmosphere.

**Other signatures of life**

There are a number of trace biogenic gases on Earth that are not detectable remotely, today or in the distant past, largely due to rapid destruction in the atmosphere. Examples include dimethyl sulfide (DMS), which is produced by marine photosynthesizers (Domagal-Goldman et al. 2011); isoprene, which is primarily produced by deciduous trees but is common across the tree of life (Zhan et al. 2021); and methyl- chloride or bromide ($CH_3Cl$ and $CH_3Br$, respectively), which are widely produced by bacteria, algae, and fungi as a detoxification strategy in a variety of marine and terrestrial environments (Leung et al. 2022). On other worlds, especially those orbiting M-dwarf host stars with lower UV fluxes, these species may accumulate to higher, remotely detectable levels (e.g. Segura et al. 2005). Investigating the detectability of trace biogenic gases and identifying novel biosignature gases remains an active area of research (Seager et al. 2016).

Absorption features arising from biogenic gases are not the only signal of life that we may be able to remotely detect. We may also be able to directly identify biological interaction with light. Most notably, absorption of visible photons and enhanced in the near-IR by the photosynthetic pigment chlorophyll is detectable in satellite observations of Earth from space and in observations of earthshine (Seager et al. 2005). This signal is largely the consequence of land vegetation today, but similar signals may arise from lichen or photosynthetic marine microbes on planets more similar to early Earth (O'Malley-James and Kaltenegger 2019). The manifestation of these "surface biosignatures" may vary on shorter timescales as well. Photosynthesis, whether by land plants or microbes, varies seasonally on Earth as light, temperature, and nutrient conditions evolve.

This seasonal variability also affects the production/consumption of gases that interact with life (Olson et al. 2018). Atmospheric $O_2$, $CH_4$, and $N_2O$ all oscillate seasonally as photosynthesis waxes and wanes. Carbon dioxide is not itself a biosignature, but it is fixed into biomass by photosynthesis and released to the environment during respiration. As a result, CO2 levels vary seasonally, and these oscillations are biogenic in origin. Temporal variability in atmospheric composition could thus reveal the presence of exoplanet life. The detectability of this biosignature depends on both the magnitude of the seasonal change as well as the average abundance of the gas, which affects the sensitivity of spectral features to changes in concentration. The detectability of Earth's seasonality has thus changed through time as the biosphere and atmospheric composition has co-evolved (Olson et al. 2018). For example, $O_2$ seasonality is negligible compared to a background of 21% by volume today, but $O_2$ seasonality could have been more dramatic in the past when $O_2$ levels were lower. The expression of seasonality may also differ among Earth-like worlds with differing obliquities or eccentricities (Jernigan et al. 2023).

Unlike $CH_4$ and $N_2O$, surface biosignatures and seasonality will not be accessible in transmission spectra with JWST. Access to surface biosignatures require reflected light spectroscopy with a direct imaging telescope, such as NASA's Habitable Worlds Observatory. As mentioned above, surface biosignatures like the red edge of vegetation are detectable in the visible and near-IR. Seasonality is also most readily characterized with direct imaging, but may be accessible in both reflected visible light and emitted light in the mid-IR with a telescope such as the Large Interferometer for Exoplanets (LIFE) mission concept (Quanz et al. 2022; Mettler et al. 2023). Consideration of how these phenomena have varied through Earth history and how their detectability may differ under a diversity of stellar, orbital, and planetary scenarios remains an exciting opportunity for future work.

## CONCLUSIONS

We conclude by summarizing the most important lessons that the early Earth can teach us for our quest of life on other planets:

(1) Earth's atmosphere was not accreted from the solar nebular in its current form; it is the product of volcanic outgassing, ingassing, loss of light volatiles to space and biological metabolisms. The redox state of outgassing volatiles has likely been oxic throughout Earth's history (*i.e.*, dominated by $N_2$, $H_2O$, $CO_2$ and $SO_2$ as opposed to $NH_3$, $H_2$, CO, $CH_4$ and $H_2S$), which carries implications for the utility of certain gases as biosignatures. Ingassing, *i.e.* recycling of volatiles back into the mantle, is not dependent on the establishment of modern-style plate tectonics. Biological gas production is dependent on metabolic innovations but also on the supply of nutrients, which ties biological activity to abiotic supplies of substrates. General trends in atmospheric evolution are summarized in Fig. 13.

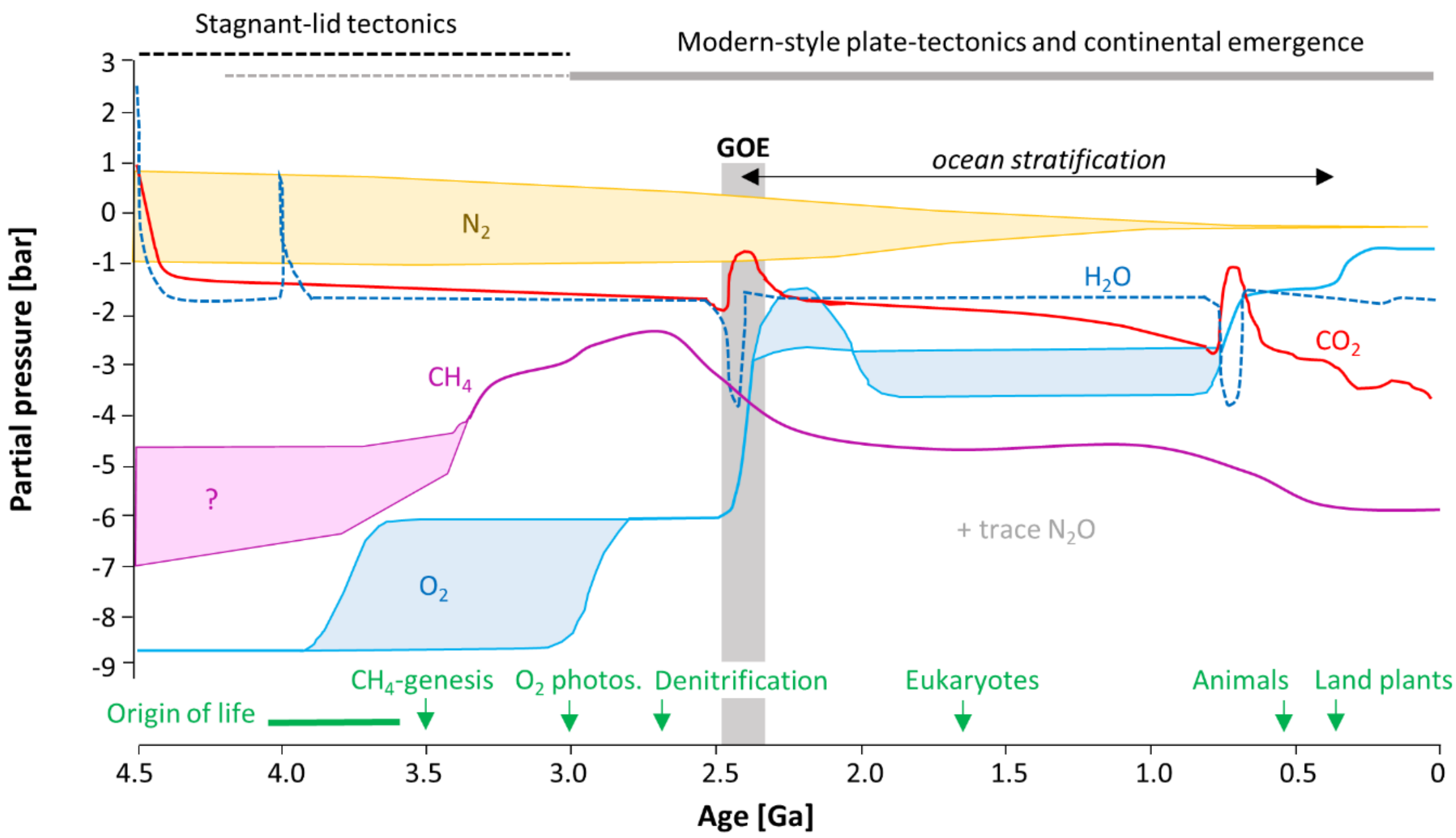

***Figure 13:*** *Trends in the abundances of major atmospheric gases over Earth's history (modified after Goldblatt 2017), along with key biological innovations (annotations in green at the bottom) and tectonic regimes (black and gray bars at the top). See text for references.*

(2) The record of life is fragmented by loss and alteration of the rock record, but remaining evidence dates back to 3.5 Ga and possibly 3.7 Ga. Life may thus have originated on a water world where continents were largely submerged and land exposure was perhaps limited to volcanic islands. The tectonic regime may have resembled a stagnant lid, akin to Venus or Mars very early in Earth history, but subduction commenced by at least 3.7 Ga and modern-style plate tectonics was in operation by c. 3 Ga. Abundant volcanism and hydrothermal activity likely generated large fluxes of $CO_2$ and $H_2$, which are substrates for biological methanogenesis. Biogenic $CH_4$ may have been abundant enough to be remotely detectable at that time. The predominance of (ultra-)mafic volcanic rocks led to elevated concentrations of Fe, Ni and Co in seawater, which was fully anoxic in the early Archean.

(3) Oxygenic photosynthesis likely evolved in the early to mid-Archean; however, due to continued input of reductants and perhaps nutrient limitation, biogenic $O_2$ was not able to accumulate in the atmosphere until 2.4 Ga (the Great Oxidation Event, GOE) and therefore not detectable remotely. Around the time when oxygenic photosynthesis emerged, the rock record shows evidence for widespread modern-style plate tectonics and continental emergence. It is conceivable that the provision of large continental shelves created new habitat space for biological innovation.

(4) From the GOE onwards, the Earth looked tectonically similar to today; however, the ocean was redox stratified with an oxygenated surface layer above anoxic bottom waters. Atmospheric $pO_2$ was likely intermediate between Archean and modern values. Anoxia was buffered by a continued supply of ferrous iron from hydrothermal sources. Seawater sulfate levels increased, which may have reduced fluxes of $CH_4$ into the atmosphere and rendered Earth's biosphere overall less detectable during its 'middle age'. A potentially important biosignature at this time may have been $N_2O$, whose production would have increased with increasing nitrate levels in surface waters and high potential for biological denitrification to $N_2O$ along the redox-cline.

It is important to note that much of these inferences rest on assumptions about total atmospheric pressure, global average surface temperature, and the major ion chemistry of seawater over time. These parameters are very difficult to constrain from existing geochemical proxies. The early Earth can therefore provide only limited insights into biosignatures on planets with widely different boundary conditions. However, if nothing else, the Earth teaches us that the establishment, maintenance and detection of life on any planet is contingent upon a favorable interplay between geological and biological processes.

## ACKNOWLEDGEMENTS

We thank Eddie Schwieterman and Martin van Kranendonk for helpful comments that improved the manuscript. We also thank Clint Scott for constructive feedback on an earlier version. The editors of this issue are thanked for the invitation to contribute. EES was financially supported by a NERC Frontiers grant (NE/V010824/1) and a Leverhulme Trust grant (RPG-2022-313) during this project. SLO acknowledges support from NASA Exobiology, NASA Habitable Worlds, NASA ICAR, and the Heising-Simons Foundation. EM was supported financially by the Energy Resources Program of the U.S. Geological Survey. Any use of trade, firm, or product names is for descriptive purposes only and does not imply endorsement by the U.S. Government. In order to meet institutional and research funder open access requirements, any accepted manuscript arising shall be open access under a Creative Commons Attribution (CC BY) reuse licence with zero embargo.